\documentclass[1p,authoryear]{elsarticle}

\usepackage{amsmath}
\usepackage{amssymb}
\usepackage{amsthm}
\usepackage{amsfonts}
\usepackage{dsfont}

\usepackage{color}
\usepackage{bookmark}

\usepackage{enumerate}

\usepackage{accents}

\graphicspath{{./fig/}}

\allowdisplaybreaks

\DeclareMathOperator*{\argmax}{arg\,max}
\DeclareMathOperator*{\argmin}{arg\,min}

\newcommand{\ubar}[1]{\underaccent{\bar}{#1}}

\newcommand{\setmap}{\mathord{\;\rightrightarrows\;}}

\newcommand{\setN}{\ensuremath{\mathcal{N}}}

\newcommand{\setV}{\ensuremath{\mathcal{V}}}
\newcommand{\setL}{\ensuremath{\mathcal{L}}}

\newcommand{\setWardper}{\ensuremath{\overset{\sim}{\mathcal{W}}}}
\newcommand{\setX}{\ensuremath{\mathcal{X}}}
\newcommand{\setS}{\ensuremath{\mathcal{S}}}
\newcommand{\setA}{\ensuremath{\mathcal{A}}}
\newcommand{\setAb}{\ensuremath{\widehat{\mathcal{A}}}}

\newcommand{\setP}{\ensuremath{\mathcal{P}}}

\newcommand{\setfeasinf}{\ensuremath{\mathcal{K}}}
\newcommand{\tcap}{\ensuremath{\theta}}
\newcommand{\tcapi}{\ensuremath{\vartheta}}

\newcommand{\tend}{\ensuremath{\eta}}

\newcommand{\setKper}{\ensuremath{\overset{\sim}{\mathcal{K}}}}
\newcommand{\setXbar}{\ensuremath{\widehat{\mathcal{X}}}}

\newcommand{\setKcirc}{\ensuremath{\overset{\circ}{\mathcal{X}}}}

\newcommand{\bfx}{\ensuremath{\mathbf{x}}}

\newcommand{\tbfx}{\ensuremath{\tilde{\mathbf{x}}}}
\newcommand{\tbfy}{\ensuremath{\tilde{\mathbf{y}}}}

\newcommand{\tphi}{\ensuremath{\tilde{\phi}}}
\newcommand{\tV}{\ensuremath{V}}
\newcommand{\bfy}{\ensuremath{\mathbf{y}}}

\newcommand{\bfw}{\ensuremath{\pmb{\omega}}}

\newcommand{\Jinf}{\ensuremath{J^{\infty}}}

\newcommand{\shift}{\ensuremath{\mathfrak{S}}}
\newcommand{\better}{\ensuremath{\mathfrak{M}}}
\newcommand{\rhmap}{\ensuremath{\mathfrak{R}}}
\newcommand{\trhmap}{\ensuremath{\tilde{\mathfrak{R}}}}

\newcommand{\teta}{\ensuremath{\psi}}%

\newcommand{\setF}{\ensuremath{\mathcal{F}}}

\newcommand{\pa}{\bar{p}} 
\newcommand{\pda}{\ubar{p}} 
\newcommand{\ta}{\bar{\tau}} 
\newcommand{\tda}{\ubar{\tau}} 

\newtheorem{thm}{Theorem}

\newtheorem{lem}[thm]{Lemma}
\newtheorem{prop}[thm]{Proposition}
\newtheorem{defn}[thm]{Definition}
\newtheorem{assum}[thm]{Assumption}

\newtheorem{rem}[thm]{Remark}

\begin{document}

\begin{frontmatter}

	\title{A framework for receding-horizon control in infinite-horizon aggregative games \tnoteref{t1}\tnoteref{t2}}
    \tnotetext[t1]{Declarations of interest: none.}
    \tnotetext[t2]{This work was supported by the EPSRC under Grant ``HubNet: Research Leadership and Networking for Energy Networks (extension)'' [ref.~EP/N030028/1] and by the
    	Leverhulme Trust under grant ECF-2016-394.}
    \address[rvt]{Electrical and Electronic Engineering Department, Imperial College London, London, UK}
    \cortext[cor1]{Corresponding author. Email: filiberto.fele@eng.ox.ac.uk}
	
	\author[rvt]{Filiberto Fele\corref{cor1}}
    \author[rvt]{Antonio De Paola}
    \author[rvt]{David Angeli}
    \author[rvt]{Goran Strbac}
	
	
\begin{abstract}
A novel modelling framework is proposed for the analysis of aggregative games on an infinite-time horizon, assuming that players are subject to heterogeneous periodic constraints. A new aggregative equilibrium notion is presented and the strategic behaviour of the agents is analysed under a receding horizon paradigm. The evolution of the strategies predicted and implemented by the players over time is modelled through a discrete-time multi-valued dynamical system. By considering Lyapunov stability notions and applying limit and invariance results for set-valued correspondences, necessary conditions are derived for convergence of a receding horizon map to a periodic equilibrium of the aggregative game. This result is achieved for any (feasible) initial condition, thus ensuring implicit adaptivity of the proposed control framework to real-time variations in the number and parameters of players. Design and implementation of the proposed control strategy are discussed and an example of distributed control for data routing is presented, evaluating its performance in simulation.
\end{abstract}
	
	
\end{frontmatter}



\section{Introduction}
This work studies the interactions between self-interested agents that autonomously pursue their individual benefit. It is supposed that the pay-off of the single agent is exclusively a function of its strategy and of the overall population behaviour. These scenarios are usually described as aggregative games~\citep{Jensen2010,KUKUSHKIN2004}, which have been recently applied in multiple contexts, including economics, power systems and transportation networks. 
Most of the studies in this area have focused on characterizing the game equilibria and devising distributed control strategies for the agents' coordination that ensure convergence and (possibly) global optimality.

The proposed analyses have mostly considered a static setting, with all the agents operating over the same finite-time horizon. This modelling framework does not fully capture realistic scenarios where the constraints and preferences of the agents, and possibly the size of the population, are expected to vary over time.
Furthermore, due to the cyclic nature of most economic, industrial, and social processes, a periodic operation can be an essential factor in pursuing optimal economical performance~\citep{AngeliEtAl2012TAC,LimonEtAl2016TACperiodic,MullerGrune2016,ZanonEtAl2013CDC,HuangEtAl2011}. However, a static framework cannot account for multiple agents that repeatedly perform heterogeneous tasks over time, as this cannot be meaningfully incorporated in a single limited time frame. To the best of the authors' knowledge, this is the first paper that tries to tackle these current limitations in the literature, extending the analysis of aggregative games to a periodic infinite-time horizon, proposing a receding horizon scheme for the distributed coordination of the agents and deriving analytical conditions for its convergence to equilibrium.

The remaining part of this introduction contains an overview on the state of the art for aggregative games and receding horizon control schemes. In addition, it summarizes the main contributions of the paper. Section \ref{sec:game_def} details the chosen modelling framework for infinite-horizon aggregative games with periodic constraints and characterizes the associated equilibria. Section \ref{sec_rhgame} presents the proposed receding horizon scheme and derives sufficient conditions for feasibility and convergence to equilibrium of better-response coordination algorithms. Finally, Section \ref{sec_exgame} presents a possible application of the proposed scheme to a data routing problem and Section \ref{sec:conc} contains some conclusive remarks.

\subsection{Relevant work - Aggregative games and distributed coordination schemes}
Noncooperative Nash games in which the objective function of a single player depends exclusively on its strategy and on some aggregation of all players' strategies have been considered in multiple papers. Although this aggregation tipically reduces to some linear function of the population strategy, much broader definitions are allowed~\citep{Jensen2010}.
In addition to theoretical works that investigate existence and uniqueness of Nash equilibria in aggregative games~\citep{Martimort2011,KUKUSHKIN2004,Dindos}, many applications and studies have been proposed in the area of economics~\citep{Novshek1985,Cornes2005}, communication networks~\citep{ALTMANEtAl2006,Basar2007}, network congestion~\citep{Alpcan2005,BarreraGarcia2015,GentileEtAl2017} and power systems~\citep{ChenEtAl2014,MaEtAl2013,DePaolaEtAl2017TSM}. Several studies have recently analysed aggregative games within the context of mean field theory~\citep{Huang2007,Lasry2007}. Research has not only focused on theoretical characterizations of the equilibrium as the population size tends to infinity, but it has also considered the design of decentralized control schemes~\citep{Nourian2013,Bauso2013,GrammaticoEtAl2016TAC}, with specific applications in several engineering-related areas~\citep{DjehicheEtAl2017}, specifically power grids~\citep{Bagagiolo2014,DePaola2016MFG}, crowd dynamics~\citep{LACHAPELLE02,AurellDjehiche2018} and economics~\citep{LACHAPELLE01}. Further research has considered the related framework of population games, characterizing the evolutionary dynamics of infinitely large collections of interacting agents~\citep{Sandholm2010PopulBook,FoxShamma2013,QuijanoEtAl2017CSM}.

In the context of aggregative games, increasing interest has been directed towards the development of distributed coordination mechanisms with suitable convergence properties.
Specific classes of games show an intrinsic convex structure which facilitates the design and analysis of noncooperative response schemes~\cite{MardenEtAl2009,HofbauerSandholm2009,FoxShamma2013,BELGIOIOSO2017}. The well-established convergence properties of these games are exploited, for example, by \citet{CandoganEtAl2013} for characterizing the limiting behaviour of general Nash games in terms of their distance from a closer potential game.
Different individual improvement paths have been considered in the literature, ranging from best-response~\citep{KUKUSHKIN2004,Jensen2010} to  other more general (possibly stochastic) strategy revision trajectories~\citep{Dindos,KUKUSHKIN2010,Lahkar2017,PovedaEtAl2017}. Other standard techniques include gradient-based schemes \citep{GrammaticoEtAl2016TAC} and variational inequality formulations~\citep{ScutariEtAl2010,BelgioiosoGrammatico2017CSL}. Current research is also focusing on the problem of coupling constraints~\citep{Grammatico2017TAC}.

Two main architectures for the implementation of these mechanisms can be distinguished. In one case, agents iteratively modify their strategy in response to an updated aggregated signal~\citep{Koshal2010,GanEtAl2013TPS,DePaolaEtAl2017CDC}. An alternative setup with a higher degree of decentralization---suitable whenever an aggregate signal broadcast is not available to the agents---adopts consensus-based techniques.
In this case, the problem is addressed through the more general set up of network games, which include an underlying game topology and characterize the agents' interactions through a graph~\citep{KoshalEtAl2016,PariseOzdaglar2017}. Each agent independently modifies its strategy (either synchronously or asynchronously) according to an estimate of the aggregate signal, based on local information exchange~\citep{GharesifardEtAl2016TAC,YeHu2017,Grammatico2017CNS}.

\subsection{Relevant work - Dynamic environment and receding horizon control}
Most of the aforementioned works on distributed coordination of the players in an aggregative game consider a finite-time horizon and a static set up. To the best of our knowledge, there is a limited number of studies that have expanded this modelling framework in order to consider a dynamic environment or explicitly account for cyclic operation.
In economics, \citet{HaurieRoche1994} define a \emph{turnpike improvement algorithm}, where the problem of market adaptation to random demand variations is addressed by tracking piecewise open-loop (infinite-horizon) Nash equilibria. More recently, in the area of power systems, \citet{GanEtAl2013TPS} have proposed a real-time distributed mechanism for charging coordination of electric vehicles, where the participation of each player to the coordination scheme can be intermittent. A different approach is presented by \citet{Song2014}, who envision a repeated game framework for the calculation of nonstationary game solutions in a demand-side management problem, allowing to partition costs among the different players on a longer time scale.

Perhaps the most suitable tool for coping with system uncertainties is the receding horizon technique, which has found broad application in engineering through model predictive control (MPC) architectures~\citep{MAYNE2014,KopanosPistikopoulos2014}.
Strongly stimulated by the increase in complexity of modern (infrastructural) systems, distributed architectures have been regarded as the natural alternative to centralized schemes.
For this reason ---especially when addressing large-population noncooperative settings--- applications of the game-theoretical framework on (model predictive) control schemes have considered noncooperative mechanisms as a means to devise distributed control laws~\citep{Scattolini2009,LiMarden2013,CHRISTOFIDES2013}. %
Starting from the work of~\citet{VANDENBROEK2002} on receding horizon solutions for a linear quadratic game, 
several distributed MPC applications of Nash games have been proposed, amongst others, for robotic formation control~\citep{Gu2008CST}, water distribution networks~\citep{RamirezJaimeEtAl2016ACC,GrossoEtAl2017}, freeway traffic control~\citep{PisarskiCanudas2016}, and economic process optimization~\citep{LeeAngeli2014}.

\subsection{Novel contributions}
This paper aims to combine the two lines of research presented above, considering aggregative games on an infinite-time horizon and proposing a novel receding-horizon scheme for the coordination of the players. In this respect, the main contributions are the following:
\begin{itemize}
	\item Formulation of a novel aggregative equilibrium concept, considering an infinite-time horizon and heterogeneous periodic constraints for the individual players (with same period but different offset). This formulation allows to account for realistic scenarios where agents need to periodically perform a task, and the incurred cost is coupled with the task scheduling of the rest of the population.
	\item New modelling framework that provides a combined description of the players' strategy update and of the receding horizon mechanism. This is obtained through a discrete-time set-valued dynamical system that returns, at each time instant, the current strategies over the predicted time horizon and the actually implemented strategy. 
	\item Application of Lyapunov stability tools and invariance results for set-valued dynamics to derive general sufficient conditions for feasibility and equilibrium convergence of receding-horizon schemes. These results are independent of the (feasible) initial conditions, thus ensuring the capability of responding to unplanned perturbations (e.g. players enter/exit the game, change their preferences) by converging to the new associated infinite-horizon equilibrium.
	\item Design of ad-hoc distributed control strategy and simulative performance evaluation for the specific problem of data routing, with quantitative assessments of convergence and robustness properties of the proposed control algorithm.
\end{itemize}

\paragraph{Notation}
Let $x\in\mathbb{R}^N$: given $i\in\setN$, $x_{-i}$ designates the vector $(x_j)_{j\in\setN\setminus\{i\}}$.
When not explicitly denoted, vectors defined in the document are intended as column vectors.
For a generic function $g$, we denote by $g^{(n)}$ the $n$-fold iterated function, i.e.,
$g^{(n)}\triangleq g\circ\ldots \circ g$, where $g^{(0)}$ is the identity function. 
We denote the \emph{restriction} operator as $(\cdot)_{\mathcal{W}}$. When applied to a generic signal $x: \mathcal{V}\rightarrow \mathcal{Y}$, it will return its restriction to the domain $\mathcal{W}\subseteq\mathcal{V}$:
\begin{equation*}
x_{\mathcal{W}} \triangleq z: \mathcal{W}\rightarrow\mathcal{Y}:\; z(\tau) = x(\tau),\, \forall \tau\in\mathcal{W}.
\end{equation*}
This notation is extended to sets $\mathcal{K}$ of signals $x: \mathcal{V}\rightarrow \mathcal{Y}$:
$$
\mathcal{K}_{\mathcal{W}} \triangleq \left \{ z :\mathcal{W}\rightarrow\mathcal{Y}:\;\exists x\in \mathcal{K},\, z(\tau) = x(\tau) \, \forall \tau \in \mathcal{W} \right \}.
$$
In the specific case where $\mathcal{W}=\{kT+\theta,\dots,(k+1)T+\theta-1\}$, the notations $x_{\langle \theta,k \rangle}$ and $\mathcal{K}_{\langle \theta,k \rangle}$ are equivalently used for $x_{\mathcal{W}}$ and $\mathcal{K}_{\mathcal{W}}$, respectively.
\par

\section{Infinite-time Horizon Aggregative Game}
\label{sec:game_def}
A population $\mathcal{N}=\left \{1, \dots, N \right \}$ of rational players is considered, denoting by $\setX_i$ the action space of player $i$. It is assumed that the players operate on an infinite-time horizon and each player must determine a certain strategy $x_i :\, \mathbb{N}\rightarrow \setX_i$ within the set of feasible strategies $\setfeasinf_i$. 
The global action space and feasible strategy set of the game are the Cartesian products $\setX = \prod_{i\in\setN}\setX_i$ and $\setfeasinf = \prod_{i\in\setN}\setfeasinf_i$, respectively.
An aggregative setup is analyzed: the stage cost of player $i$ at time $t$ depends on its current action $x_i(t)\in \setX_i$ and on some aggregation $\sigma(x(t))$, of the current actions $x(t)\in \setX$ of all players, with $\sigma: \setX \to \mathbb{R}^m$ and $m\geq 1$. The cost function $J^{\infty}$ for each player $i\in \mathcal{N}$ can therefore be expressed as:
\begin{equation}
\Jinf_i(x_i,x) \triangleq \limsup_{t\rightarrow\infty}\frac{1}{t} \sum_{\tau=0}^{t-1} \pi\left(\sigma(x(\tau)), \, x_i(\tau) \right)
\label{eq_indcost} 
\end{equation}
The proposed formulation of $\Jinf_i$ corresponds to an \textit{infinite-horizon undiscounted average cost}. As mentioned in the Introduction, we are interested in analyzing aggregative games where the cost functions and constraints of each player are not limited to a time interval of finite length. For this reason, the cost function is expressed as the limit sums of the stage cost $\pi$. In order to operate with bounded quantities and ensure that $\Jinf_i$ is always well defined, the cost is averaged over the whole time interval and the limit superior operation is considered. The choice of not introducing any discount factor is motivated by the underlying periodic framework. Note that, as a result of the average and limit operations in (\ref{eq_indcost}), the proposed cost function neglects the transients and represents instead the steady-state behaviour of the players. 

The following definition can now be provided for the considered game:
\begin{defn} %
\label{def_gamestatic}
The infinite-time horizon aggregative game is defined as the tuple $\langle \setN, (\setfeasinf_i)_{i\in\setN}, (\Jinf_i)_{i\in\setN} \rangle$, where
$\setN$ is the set of players, $\setfeasinf_i$ is the set of feasible \emph{strategies} of player $i\in\setN$ and $\Jinf_i :\, \setfeasinf_i\times\setfeasinf\to\mathbb{R}$ in (\ref{eq_indcost}) is the cost function of player $i\in\setN$. %
\end{defn}

We wish to point out that the proposed formulation can naturally be extended, considering an additional periodic cost term $g_i$ in (\ref{eq_indcost}) that is different in general for each player. Under continuity and convexity of $g_i$, the analysis presented in the rest of the paper remains valid. A detailed study of this more general case will be presented in a future work.

\subsection{Equilibrium Characterization}
The theoretical analysis and the study of equilibria for the infinite-time horizon aggregative game introduced in Definition \ref{def_gamestatic} are carried out under the following assumptions:
\begin{assum}
	\label{assum:sigma}
The functions $\sigma:\setX \rightarrow \mathbb{R}^m$  and  $\pi : \mathbb{R}^m \times \setX_i \rightarrow \mathbb{R}$ are continuous.
\end{assum}
\begin{assum}
	\label{assum:pi}
	For any $\bar{\sigma}\in\mathbb{R}^m$, the function $\pi(\bar{\sigma},\cdot)$ is convex on $\setX_i$.
\end{assum}

It is now possible to introduce the equilibrium concept that will be considered as the main design objective in the rest of the paper:
\begin{defn}
\label{def:WE}
Let $\mathcal{W} \subseteq \setfeasinf$ denote the set of infinite-horizon aggregative equilibria, defined as follows:
\begin{equation}
\label{eq:WE}
\mathcal{W} \triangleq \left \{ x^*=(x^*_i)_{i\in\setN}\in\setfeasinf \, :  \, \Jinf_i(x^*_i,x^*) \leq \Jinf_i(x_i,x^*) \quad \forall x_i\in\setfeasinf_i,\, \forall  i\in\setN  \right \}
\end{equation}
\end{defn}
We wish to emphasize that, under the proposed equilibrium notion, the aggregative term $\sigma(x)$ in (\ref{eq_indcost}) is considered unaffected by unilateral strategy changes by player $i$. This 
choice represents a reasonable approximation of the classic Nash equilibrium notion for large numbers of players. The two equilibrium concepts become equivalent when the number of players goes to infinity.\footnote{In line with the definition of Wardrop equilibrium.}

\subsection{Heterogeneous Periodic Constraints}
\label{sec_periodic}
\begin{figure}
	\centering
	\def\svgwidth{1.5\columnwidth}
	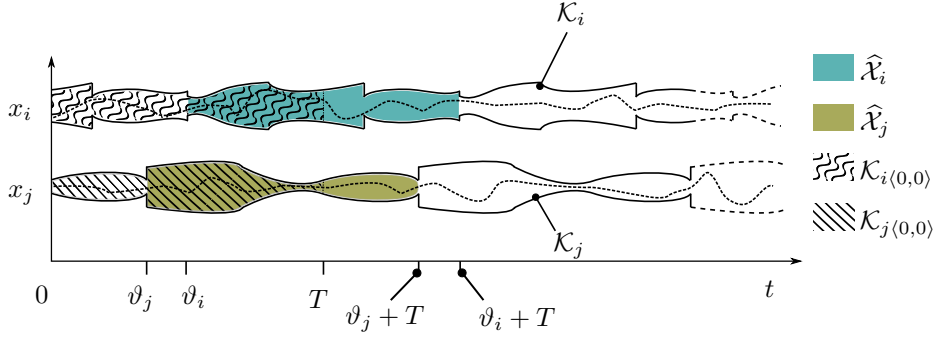
	\caption{Graphical example of periodic constraints for two players $i$ and $j$.
	}
	\label{fig_schema_period}
\end{figure}

The infinite-time horizon aggregative game introduced in Definition \ref{def_gamestatic} is now analysed in a periodic setting. Specifically, it is assumed that the strategy $x_i$ of the individual player $i\in\setN$ must periodically fulfil some constraints, characterized by the compact and convex set $\setXbar_i$.
Period and phase (\emph{offset}) of the constraints are denoted by $T_i=T$ (equal for all players) and $\tcapi_i$, respectively. Recalling that $x_{ \langle \theta, k \rangle}$ denotes the restriction of $x(\cdot)$ on the interval $\{kT+\theta,\dots, k(T+1)+\theta -1\}$, the set $\setfeasinf_i$ of feasible strategies for player $i$ can be defined as:
\begin{equation}
\setfeasinf_i \triangleq \{x_i\colon \mathbb{N}\to\setX_i:\, x_{i\langle\tcapi_i,k \rangle}\in\setXbar_i,\, \forall k\in\mathbb{N}\}.
\label{eq_setfeasinf}
\end{equation}
From (\ref{eq_setfeasinf}), the strategy of the individual player $i$ on periodic intervals of length $T$ must belong to a certain feasibility set $\setXbar_i$. The phase parameter $\tcapi_i$ of player $i$ determines the offset with respect to $t=0$ of this periodic time interval, which will correspond to $\{kT+\tcapi_i, \dots, k(T+1)+\tcapi_i -1 \}$ with $k\in \mathbb{N}$.
An example in such sense is provided in Figure \ref{fig_schema_period}, where it can be seen how the feasibility sets $\setfeasinf_i$ and $\setfeasinf_j$ for the generic players $i,j \in \mathcal{N}$ are constituted by the periodic repetition of the sets $\setXbar_i$ and $\setXbar_j$, respectively, whose offset with respect to $t=0$ is equal to $\tcapi_i$ and $\tcapi_j$. The shaded areas $\setfeasinf_{i\langle0,0\rangle}$ and $\setfeasinf_{j\langle0,0\rangle}$ represent the restrictions of the sets $\setfeasinf_{i}$ and $\setfeasinf_{j}$ on the interval $\{ kT+\theta,\dots,(k+1)T+\theta-1\}$ for $k=0$ and $\theta=0$.

The proposed characterization of strategy feasibility can model for example multi-agent problems (e.g. charging of electric vehicles) where the individual agents need to periodically perform a certain task over time, with overlapping (and possibly unaligned) availability time windows of different agents.
\begin{rem}
The more general case of individual agents $i$ with constraints of different period $T_i$ can be accommodated in the proposed analysis by setting $T$ as the least common multiple of $\{T_i\}_{i\in\setN}$. 
\end{rem}
In order to characterize the equilibria of the game in Definition \ref{def_gamestatic} under the proposed periodic framework, we focus on the set of periodic strategies $\setKper\subseteq \mathcal{K}$, defined as follows:
\begin{equation}
\label{eq:Ktilde}
\setKper:= \left \{  x \in \mathcal{K} : x_{\langle \theta, k_1 \rangle} = x_{\langle \theta, k_2 \rangle}, \quad \forall k_1,k_2 \in \mathbb{N}, \quad \forall \theta \in \{ 0, \dots, T-1 \} \right \}
\end{equation}
Moreover, the  functional $J_i$ is derived from (\ref{eq_indcost}) in order to evaluate the cost for player $i$ on a single period of length $T$:
\begin{equation} %
J_i(x_i,x) = \sum_{t = 0}^{T-1}\pi\left(\sigma(x(t)), x_i(t)\right).
\label{eq_finitehor_cost}
\end{equation}
The set $\setWardper$ of periodic aggregative equilibria can now be introduced:
\begin{equation}
\label{eq:Wtilde}
\setWardper = \left \{ x^* \in \setKper : J_i(x^*_{i\langle 0,k \rangle},x^*_{\langle 0,k \rangle}) \leq J_i(z_i,x^*_{\langle 0,k \rangle}), \,\,  \,\forall z_i \in \setKper_{i\langle 0,k \rangle}, \, \forall i \in \mathcal{N}, \, \forall k\in \mathbb{N} \right \}.
\end{equation}
Note that $\setWardper$ contains all periodic strategies that fulfil the equilibrium conditions locally, i.e. on each individual time period.
It is now verified that such strategies are also equilibria on the infinite time-horizon:
\begin{lem}
	\label{lem_setaggreq}
	Given $\setWardper$ in (\ref{eq:Wtilde}) and the set of infinite-time horizon aggregative equilibria $\mathcal{W}$ in (\ref{eq:WE}), it holds:
	\begin{equation}
	\label{lem:WE}
	\varnothing \subset \setWardper \subseteq \mathcal{W}.
	\end{equation}
\end{lem}
\begin{proof}
\textit{Non-emptiness:} The strict set inclusion in (\ref{lem:WE}) is initially demonstrated, verifying that $\setWardper \neq \varnothing$. To this end, we introduce the set $\setKcirc = \prod_{i\in\setN} \setKcirc_i$, with $\setKcirc_i$ defined as follows:
\begin{equation}
\setKcirc_i \triangleq \left \{y_i \in \setX_i^T : \exists z_i \in \setXbar_i, y_i(\bmod(t+\tcapi_i,T))=z_i(t), \forall t\in\{ 0, \dots, T-1 \} \right \}
\label{eq_setcirc}
\end{equation}
where $\setX_i^T$ can be interpreted as a set of signals over a time horizon of length $T$.
One can verify that $\setKcirc$ is convex and compact since the same properties hold for $\setXbar_i$ in (\ref{eq_setfeasinf}).
Consider now the set-valued mapping $\mathfrak{F}=(\mathfrak{F}_i)_{i\in\setN}\colon \setKcirc\setmap\setKcirc$, with $\mathfrak{F}_i(y)$ defined as follows:
\begin{equation*}
\mathfrak{F}_i(y) = \argmin_{w_i\in\setKcirc_i} J_i(w_i,y),\quad \forall i\in\setN.
\end{equation*}
It follows from Assumption \ref{assum:sigma} and \ref{assum:pi} that $J_i(w_i,y)$ in (\ref{eq_finitehor_cost}) is continuous (in both variables) and convex in $w_i$. Given the convexity of $\setKcirc_i$, it follows from \cite[Theorem~9.17]{RangarajanOptTheoryBook1996}  that $\mathfrak{F}_i(y)$ is a convex-valued upper semi-continuous correspondence and therefore it has closed graph.
From the Kakutani fixed-point theorem \citep{Aliprantis06,Kakutani1941} there exists $w^*\in\setKcirc$ such that $w^*\in\mathfrak{F}(w^*)$. Considered $x^*$ with $x^*_{\langle 0,k \rangle} = w^*$ for all $k\in \mathbb{N}$, we have that $x^*\in \setWardper$ in (\ref{eq:Wtilde}), thus proving $\setWardper \neq \varnothing$.
\\
\textit{Aggregative equilibrium:} The second set inclusion in (\ref{lem:WE}) is now checked, verifying that each element $x^* \in \setWardper$ also belongs to the set of infinite-horizon aggregative equilibria $\mathcal{W}$. In this respect it is sufficient to show that, for any feasible $x\in \mathcal{K}$, we have:
\begin{multline}\label{eq_we}
\min_{x_i\in\setfeasinf_i}\Jinf(x_i,x^*) \triangleq \min_{x_i\in\setfeasinf_i}\limsup_{t\rightarrow\infty} \frac{1}{t}\sum_{\tau=0}^{t-1}\pi(\sigma(x^*(\tau)),x_i(\tau) ) \\
\geq \min_{x_i\in\setfeasinf_i}\limsup_{k\rightarrow\infty} \frac{1}{kT} \sum_{\tau=0}^{kT-1}\pi\left(\sigma(x^*(\tau)),x_i(\tau) \right) \qquad \qquad \qquad \qquad \\
= \min_{x_i\in\setfeasinf_i}\limsup_{k\rightarrow\infty} \frac{1}{kT} \sum_{\tau=\tcapi_i}^{\tcapi_i+kT-1}\pi\left(\sigma\left(x^*(\tau)\right),x_i(\tau)\right) \qquad  \qquad   \\
\geq \limsup_{k\rightarrow\infty} \frac{1}{kT} \min_{x_i\in\setfeasinf_i}\sum_{\tau=\tcapi_i}^{\tcapi_i+kT-1}\pi\left(\sigma\left(x^*(\tau)\right),x_i(\tau)\right) \\ 
\overset{(\ref{eq_we}a)}{=} \limsup_{k\rightarrow\infty} \frac{1}{kT} \sum_{l=0}^{k-1}\min_{ y_i \in \setXbar_i }\sum_{\tau=\tcapi_i + lT}^{\tcapi_i+(l+1)T-1}\pi\left(\sigma\left(x^*(\tau)\right),y_i(\tau)\right)\\
\qquad \overset{(\ref{eq_we}b)}{=} \limsup_{k\rightarrow\infty} \frac{1}{kT} \sum_{l=0}^{k-1}\min_{y_i\in \setKcirc_i }\sum_{\tau=lT}^{(l+1)T-1}\pi\left(\sigma\left(x^*(\tau)\right),y_i(\tau)\right)\\
\overset{(\ref{eq_we}c)}{=}  \frac{1}{T} \sum_{\tau=0}^{T-1}\pi\left(\sigma\left(x^*(\tau)\right), x^*_{i\langle 0, 0\rangle} \right)
\\
 = \limsup_{t\rightarrow\infty}\frac{1}{t}\sum_{\tau=0}^{t-1}\pi\left(\sigma\left(x^*(\tau)\right), x_i^*(\tau)\right) = \Jinf(x_i^*,x^*) \geq \min_{x_i\in\setfeasinf_i}\Jinf(x_i,x^*).
\end{multline}
Note that (\ref{eq_we}a) is always verified, since $\mathcal{K}_i$ in (\ref{eq_setfeasinf}) is characterized by periodic constraints $\setXbar_i$ with period $T$ and phase $\tcapi_i$. As a result, the minimization can always be performed by separate optimizations on each time period. The equality (\ref{eq_we}b) follows from (\ref{eq_setcirc}) and independence of the sums of $\pi$ with respect to $\tcapi_i$ (periodicity of $x^* \in \setWardper \subseteq \setKper$). Finally (\ref{eq_we}c) follows from the inequalities in (\ref{eq:Wtilde}) for $k=0$, since $\setKcirc_i$ can alternatively be expressed as $
\setKcirc_i = \left \{ y_i : \exists x \in \setKper, x_{i\langle 0,0 \rangle} = y_i \right \}
$.
\end{proof}

\section{Receding-horizon framework}
\label{sec_rhgame}
Having introduced the infinite-time horizon aggregative game of Definition \ref{def_gamestatic}, we are now interested in designing closed-loop strategies that ensure convergence to the set of aggregative equilibria characterized in Definition \ref{def:WE}. In particular, a receding-horizon scheme with prediction horizon of length $T$ is proposed. This is modelled by a discrete-time dynamical system with state $\tbfx= [ \bfx \,\,\,\, \theta ]$, where $\bfx\in \setX^T$ (in bold notation) represents the predicted strategy of the agents over the next $T$ time instants and $\theta\in \{ 0, \dots, T-1 \}$ denotes the current phase with respect to the period $T$. The dynamical system (with initial state $\tbfx(0) = [\bfx^0,0]$) can then be described by the following state equations:
\begin{subequations}
	\label{eq:state}
	\begin{equation}
	\label{eq:state1}
	\bfx(t+1) \in \rhmap(\bfx(t),\theta(t) )
	\end{equation}
	\begin{equation}
	\label{eq:state2}
	\theta(t+1) = \bmod (\theta(t)+1,T)
	\end{equation}
\end{subequations}
and by the following output:
\begin{equation}
\label{eq:output}
y(t) = \bfx_{\left\{0\right\}}(t)
\end{equation}
where $\bfx_{\left \{ \tau \right\}} (t) \in \setX $ is used to denote the $\tau$-th (time) component of $\bfx(t)\in \setX^T$.
The output $y$ corresponds to the actual strategy played by the agents. Consistently with the envisioned receding horizon approach,  (\ref{eq:output}) imposes that, at each time $t$, the output $y(t)$ is equal to the first element of the agents' predicted strategy $\bfx(t)$ (which ranges from time $t$ to time $t+T-1$).\par
From (\ref{eq:state1}), the predicted strategy $\bfx(t+1)$ at time $t+1$ (i.e. the predicted actions over the time interval $\{t+1,\dots,t+T\}$) depends on the corresponding quantity $\bfx(t)$ at the previous time instant according to the multi-valued mapping $\rhmap$, i.e. the receding-horizon map. The mapping $\rhmap$ is characterized as the composition of two submappings:  a rotation operator $\shift$ and a strategy revision policy $\better$.
The rotation operator $\shift$ is used to generate an initial feasible strategy over the new prediction horizon $\{t+1,\dots,t+T\}$. Under the considered periodic framework, this can be obtained by simply shifting the current strategy by one time step. The preliminary result obtained with $\shift$ is then modified by the update map $\better$ according to some strategy revision policy, e.g. better response.

A formal definition of $\rhmap$ and the submappings $\shift$ and $\better$ is now provided: 
\begin{defn}[Receding-horizon map]
	\label{def_rhmap}
	The map $\rhmap: \setX^T \times  \{ 0, \dots, T-1 \} \setmap \setX^T$ is defined as the composition $\rhmap \triangleq \better \circ \shift$, where:
	\begin{itemize}
		\item (Rotation) $\shift\colon \setX^T \to \setX^T$. Given $\bfx \in{\setX^T}$ and the associated $\shift(\bfx)\in{\setX^T}$, the individual time-component $\shift_{\left \{ s \right\} }(\bfx) \in{\setX}$ can be expressed as:
		\begin{equation}
		\label{eq:Gmapping}
		\shift_{\left \{ s \right\} }(\bfx) = \bfx_{\{\mathrm{mod}(s+1,T)\}} \qquad s=0,\dots,T-1.
		\end{equation}
		\item (Update map) $\better \colon \setX^T \times  \{ 0, \dots, T-1 \} \setmap \setX^T$.
		\end{itemize}
\end{defn}
\noindent According to (\ref{eq:Gmapping}), $\shift(\bfx(t))$ corresponds to the last $T-1$ elements of the previous strategy $\bfx(t)$, i.e. $\bfx_{ \{2,\dots,T\}}(t)$ defined on $\{t+1,\dots,t+T-1\}$, concatenated with the first element $\bfx_{ \left \{ 1 \right \}}(t)$.

The dynamical system (\ref{eq:state}) can equivalently be represented in the following compact manner:
\begin{equation}
\label{eq:state_compact}
\tilde{\bfx}(t+1) = [\bfx(t+1),\theta(t+1)] = \tilde{\rhmap}(\tilde{\bfx}(t))
\end{equation}
with the multi-valued mapping $\tilde{\rhmap}:\setX^T \times \{ 0, \dots, T-1 \} \setmap  \setX^T \times \{ 0, \dots, T-1 \}$ defined as follows:
\begin{equation}
\label{eq:Rtilde}
\tilde{\rhmap}(\tbfx)=\tilde{\rhmap}([\bfx,\theta]) \triangleq \left \{ [\bfx^+,\theta^+] : \bfx^+ \in \rhmap(\bfx,\theta) , \theta^+= \bmod(\theta+1,T)\right \}.
\end{equation}

\subsection{Properties of the RH mapping}
\begin{figure}
	\centering
	\def\svgwidth{0.7\columnwidth}
	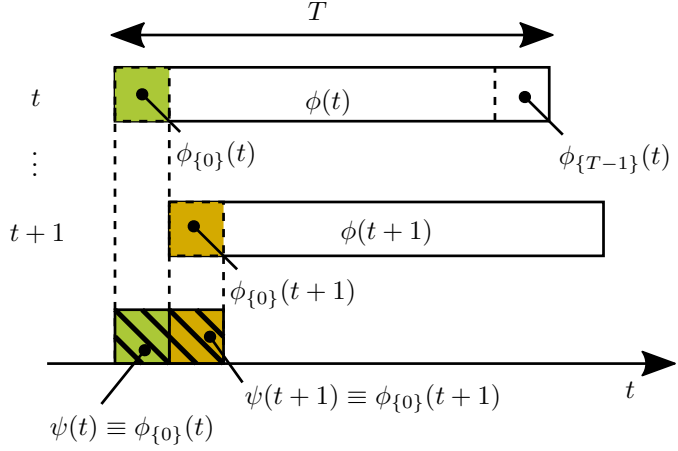
	\caption{Diagram of the receding-horizon solution.}
	\label{fig_schemaRH}
\end{figure}

To characterize the solutions of system (\ref{eq:state}), it is assumed without loss of generality that $\theta(0)=0$ and the following definition is provided:
\begin{defn}[Solution set]
	\label{def_solset}
We denote by $\Phi(\bfx^0)$ the set of solutions of system~\eqref{eq:state} with initial condition $\bfx(0) =\bfx^0$ and $\theta(0)=0$:
\begin{equation}\label{eq_solset}
\Phi(\bfx^{0}) \triangleq \{\tilde{\phi}(\cdot) = [\phi(\cdot),\Theta(\cdot)] :  \phi(0) = \bfx^0,\,
 \phi(t+1)\in \rhmap(\phi(t),\Theta(t)),\,\forall t\in\mathbb{N} \}.
\end{equation}
From~\eqref{eq:output}, the output trajectory associated to $\tilde{\phi}(\cdot)=[\phi(\cdot),\Theta(\cdot)] \in \Phi(\bfx^{0})  $ is denoted as $\teta(\cdot)$, with $\psi(t)=\phi_{\{0\}}(t)\in\setX$.\footnote{Dependence of $\teta$ from $\phi$ is not explicitly represented for compactness of notation.}
\end{defn}
Note that, since the values of the state component $\theta(t)$ exclusively depend on time, the associated solution component in (\ref{eq_solset}) has always the following expression:
\begin{equation}
\label{eq:Theta_expr}
\Theta(t) = \bmod (t,T).
\end{equation}
We recall that, under the proposed receding horizon framework, $\phi(t)\in \setX^T$ and $\teta(t)\in \setX$ represent, respectively, the planned strategy on $\{t,\dots,t+T-1\}$ and the actual strategy at time $t$ implemented by the players of the game in Definition \ref{def_gamestatic}. A graphical representation is provided in Figure \ref{fig_schemaRH}.

The convergence of the receding horizon solution to the set of periodic aggregative equilibria is demonstrated with Lyapunov arguments, considering a continuous function $g: \setX \rightarrow \mathbb{R}_+$ and a Lyapunov function $V: \setX^T \rightarrow \mathbb{R}_+$ of the following form:
\begin{equation}
\label{eq:Vdef}
V(\bfx) = \sum_{\tau=0}^{T-1} g \left ( \bfx_{\left \{ \tau \right \}} \right )
\end{equation}
To establish convergence, some fundamental properties must be fulfilled by the mapping $\better$ in Definition \ref{def_rhmap}. In addition to its continuity, it is supposed that the application of $\better$ preserves feasibility of the planned strategy. This means that, if $\bfx$ can be infinitely replicated over time to generate a feasible periodic strategy, the same must hold for $\better(\bfx,\theta)$. Furthermore, it is assumed that the application of $\better$ corresponds to a reduction of the Lyapunov function $V$.
To additionally demonstrate that convergence is towards an equilibrium of the game introduced in Definition \ref{def_gamestatic}, it is also imposed that the mapping $\better$ does not modify the planned strategy $\bfx$ if the local cost function $J_i$ in (\ref{eq_finitehor_cost}) of certain players is minimized.

All the requirements above are formally stated as follows:
\begin{assum}
	\label{assum_rhmap}
	The mapping $\better : \setX^T \times \{ 0, \dots, T-1 \}\setmap \setX^T$ in Definition \ref{def_rhmap} and the considered Lyapunov function $V:\setX^T \rightarrow \mathbb{R}_+$ fulfil the following properties:
	\begin{enumerate}[(a)]
		\item \label{assum_rhmap_a} The mapping $\better$ is upper semi-continuous.
		\item \label{assum_rhmap_b} The mapping $\better$ is feasibility-invariant: for any $\theta \in \{ 0, \dots, T-1 \}$ and $\bar{\theta}= \bmod(\theta+1,T)$, it holds
		\begin{equation}
        \label{eq:rh_map_b}
		\better(\bfx,\theta) \subseteq \setKper_{\langle \bar{\theta}, k \rangle} \qquad \forall  k \in \mathbb{N},  \, \forall \bfx\in\setKper_{\langle \bar{\theta},k \rangle}.
		\end{equation}
		\item \label{assum_rhmap_c} For some positive definite function $\rho:[0,+\infty] \rightarrow [0,+\infty]$, it holds:
        \begin{equation}
        \label{eq:rh_map_c}
		V(\bfx^+) \leq V(\bfx) - \rho(\|\bfx^+ -\bfx\|) \qquad \forall \bfx \in \setX^T , \quad  \forall \bfx^+\in\better(\bfx,\cdot)
		\end{equation}
		\item \label{assum_rhmap_d}
		For any $\theta=\{ 0, \dots, T-1 \}$, it holds:
		\begin{equation}
		\label{eq:rh_map_d}
		\bfx \in \better(\bfx,\theta) \, \Rightarrow
		\left \{
		\begin{array}{l}
				J_i(\bfx_i,\bfx) =  \displaystyle \min_{y_i\in\setXbar_i} J_i(y_i,\bfx) \qquad \forall i: \tcapi_i = \bmod(\theta+1,T)
		\\
		\better(\bfx,\theta) = \left \{ \bfx \right \}.
		\end{array}
		\right .
		\end{equation}
	\end{enumerate}
\end{assum}
Point~\eqref{assum_rhmap_b} in Assumption~\ref{assum_rhmap} ensures that the mapping $\better$ preserves the feasibility of the predicted strategy $\bfx$.
Note that $\bar{\theta}= \bmod(\theta+1,T)$ is used in (\ref{eq:rh_map_b}) as $\better$ is applied to a predicted strategy that has already been shifted to the next time interval by $\shift$.  Condition~\eqref{assum_rhmap_c} specifies that the application of the mapping $\better$ always yields nonpositive variations in the function $V$. Finally \eqref{assum_rhmap_d} ensures that, when the mapping $\better$ admits the input strategy $\bfx$ among its output values, the following conditions are verified:
\begin{itemize}
\item all players whose periodic constraints are aligned with the current prediction window are minimizing their local cost function $J_i$. From (\ref{eq_setfeasinf}), these players are characterized as all $i\in \mathcal{N}$ such that $\tcapi_i = \bmod(\theta+1,T)$, since $\better$ operates on a predicted strategy that has already been shifted forward in time by $\shift$.
\item no other strategy different from $\bfx$ is possible.
\end{itemize}

\begin{rem}
	It can be seen from (\ref{eq:rh_map_c}) and (\ref{eq:rh_map_d}) that the relation between the decrease of the value of $V$ and that of the individual costs $J_i$ is only imposed at equilibrium. As a result, the above assumptions are fulfilled by potential games but in general include a wider range of cases.
\end{rem}

\subsection{Control Design and Implementation}
Assumption \ref{assum_rhmap} provides general conditions for feasibility and convergence to equilibrium of the proposed receding horizon scheme, as demonstrated in Section \ref{eq:conv_eq_res}. These conditions also give useful indications on the design of the mapping $\better$ and the choice of the Lyapunov function $V$. One possibility in such sense is to consider sequential better-response schemes and define $\better$ as the composition of $N$ submappings $\better_1, \dots, \better_N$, with $\better(\bfx,\theta) = (\better_N \circ \dots \circ \better_1)(\bfx,\theta)$,\footnote{With a slight abuse of notation, composition is carried out only with respect to the $\bfx$ argument.} where $\better_i$ corresponds to a strategy update performed by player $i$ in order to reduce its local cost $J_i$ over the feasibility set $\setKper_{i\langle \bar{\theta}, k \rangle}$, with $ \bar{\theta} = \bmod(\theta+1,T)$. When $\better$ does not modify the planned strategy, it follows that all players $i$ with $\vartheta_i=\bar{\theta}$ cannot further improve their local cost and condition (\ref{eq:rh_map_d}) is verified by default, since $\setKper_{i\langle \bar{\theta}, k \rangle} \equiv \setXbar_i$ when $\bar{\theta} = \vartheta_i$. As a result, Assumption \ref{assum_rhmap} can be fulfilled by verifying its points (a), (b) and (c) on the individual submappings $\better_i$. Point (a) can be verified by envisioning strategy updates that are continuous with respect to some scalar function of the system state (for example prices) whereas (b) is implicitly fulfilled under the present framework, as no player will opt for an unfeasible update. Finally, in order to satisfy point (c), one can consider a Lyapunov function $V$ that is related to the total cost of the players. This choice is naturally fitting for potential games but can be extended to a larger class of problems if some limitation is imposed on the update by the individual players. An example of the proposed approach for the design of $\better$ and the choice of $V$ is presented in Section \ref{sec_exgame}  for the case of data routing over a multi-hop network.

Regarding the implementation of the receding-horizon map presented in Definition \ref{def_rhmap}, we wish to emphasize that the proposed formulation (and subsequent theoretical analysis) represent a general theoretic construct that can be used to model and accommodate multiple heterogeneous scenarios. The mapping $\better$ can be specifically designed in order to consider coordination schemes for the players that have different levels of centralization and communication requirements. In principle, centralized control, distributed signal-based coordination and decentralized consensus schemes can all be described under the proposed framework. For example, if one operates with the better-response scheme described above, the receding horizon map introduced in Definition \ref{def_rhmap} can be implemented through a bi-directional communication channel between the players and some central entity. The latter broadcasts some control signal (for example prices) to the individual player $i$, which in turn updates its strategy according to $\better_i$ and communicates its strategy change to the central entity. An updated control signal can then be calculated and the whole procedure is repeated for the other players. This kind of framework can be applied, for example, to the data routing problem presented in Section \ref{sec_exgame}.
	
\subsection{Convergence and equilibrium results}
\label{eq:conv_eq_res}
In order to consider an appropriate convergence notion and demonstrate that the state and output of system (\ref{eq:state})-(\ref{eq:output}) tend to the set of infinite-horizon aggregative equilibria, the definition of $\omega$-limit set is preliminarily introduced:
\begin{defn}[$\omega$-limit set, Definition 6.17~\cite{GoebelEtAl_HybridDynSysBOOK}]
For a solution $\tphi(\cdot) = [\phi(\cdot),\Theta(\cdot)]\in \Phi(\bfx^0)$ of system (\ref{eq:state}), the associated $\omega$-limit set $\Omega(\tphi)$  is defined as:
\begin{equation}
\Omega(\tphi)\triangleq \{\tilde{\bfw}=[\bfw,\theta] :\, \exists \{t_n\}_{n\in\mathbb{N}},\, \lim_{n\rightarrow\infty}t_n=\infty,\,\lim_{n\rightarrow\infty}\phi(t_n)=\bfw, \, \lim_{n\rightarrow\infty} \Theta(t_n)=\theta\}.
\label{eq_omegalimset}
\end{equation}
\end{defn}

Recalling expression (\ref{eq:Theta_expr}) for the solution component $\Theta(t)$ associated to the state variable $\theta$, the $\omega$-limit set can equivalently be expressed as the following partition:
$$
\Omega(\tilde{\phi}) =  \bigcup_{\theta \in \{ 0, \dots, T-1 \}} \Omega^{\theta}(\tilde{\phi}) \times \left \{\theta \right \}
$$
where each subset $\Omega^{\theta}(\tilde{\phi})$ has the following expression:
\begin{equation}
\Omega^{\theta}(\tphi)\triangleq \{\bfw:\, \exists \{t_n\}_{n\in\mathbb{N}},\, \lim_{n\rightarrow\infty}t_n=\infty,\,\lim_{n\rightarrow\infty}\phi(t_n)=\bfw, \, \lim_{n\rightarrow\infty} \Theta(t_n)=\theta\}.
\label{eq_omegalimset_tau}
\end{equation}

Some relevant properties of the $\omega$-limit set are now demonstrated:
\begin{lem}%
	\label{lem:omega_prop}
	Let Assumption~\ref{assum_rhmap}\eqref{assum_rhmap_a}-\eqref{assum_rhmap_c} hold. For any $\tilde{\phi}(\cdot) = [\phi(\cdot), \Theta(\cdot) ]\in \Phi(\bfx^0)$ and associated $\Omega(\tilde{\phi})$ it holds:
	\begin{subequations}
		\label{eq:omega_prop}
		\begin{equation}
		\label{eq:omega_prop_a}
		\lim_{t\rightarrow\infty} |\tphi(t)|_{\Omega(\tphi)} = 0,
		\end{equation}
		\begin{equation}
		\label{eq:omega_prop_b}
		\lim_{t\rightarrow\infty} \tV(\phi(t)) = \underline{V} = V(\bfw) \qquad \forall \tilde{\bfw}= [\bfw,\theta] \in \Omega(\tilde{\phi}).		
		\end{equation}
		\begin{equation}
		\label{eq:omega_prop_c}
		\rhmap(\bfw,\theta) = \left \{ \shift(\bfw)  \right\} \subseteq  \Omega^{\bar{\theta}} (\tilde{\phi})
		\qquad
		\forall  \bfw \in \Omega^{\theta}(\tilde{\phi}), \quad \forall \theta \in \{ 0, \dots, T-1 \}
		\end{equation}
		where $\bar{\theta}=\bmod(\theta+1,T)$, the mapping $\tilde{\rhmap}$ is defined in (\ref{eq:Rtilde}) and $|y|_{\mathcal{A}}$ denotes distance of $y$ from the set $\mathcal{A}$.
	\end{subequations}
\end{lem}
\begin{proof}
	\textit{Condition (\ref{eq:omega_prop_a})}:\hspace{1ex}The upper semicontinuity of $\tilde{\rhmap}$ is preliminarily demonstrated by showing that the graph $\mathbf{grf}(\trhmap)$ of $\tilde{\rhmap}$ is closed~\citep{Aliprantis06}:	
\begin{multline}
\label{eq_grafR}
\mathbf{grf}(\trhmap) \triangleq \left\{(\tbfx,\tbfy):\,
\tbfx\in\setX^T\times \{ 0, \dots, T-1 \},\, \tbfy\in\trhmap(\tbfx)\right\} \\
= \bigcup_{\theta\in\{ 0, \dots, T-1 \}} \left\{ ([\bfx,\theta],[\bfy,\bmod(\theta+1,T)]):\,
\bfx\in\setX^T,\, \bfy\in\rhmap(\bfx,\tcap)\right\}. 
\end{multline}
From Assumption~\ref{assum_rhmap}\eqref{assum_rhmap_a}, the mapping $\better(\bfx,\tcap)$ is upper semicontinuos, for any $\bfx\in\setX^T$ and $\tcap\in \{ 0, \dots, T-1 \}$. This and the continuity of $\shift$ imply upper semicontinuity of $\rhmap = \better \circ \shift$ in Definition \ref{def_rhmap} \citep{Aliprantis06}.
As a result, the graph $\mathbf{grf}(\rhmap(\bfx,\tcap)) = \left\{
(\bfx,\bfy):\, \bfx\in\setX^T,\, \bfy\in\rhmap(\bfx,\tcap)\right\}$ is closed. 
It follows from the right-hand side in (\ref{eq_grafR}) that $\mathbf{grf}(\trhmap)$ is the union of a finite number of closed sets and therefore it is also closed, thus ensuring upper semicontinuity of $\tilde{\rhmap}$ from previous considerations. From \cite{GoebelEtAl_HybridDynSysBOOK}, this result straightly implies (\ref{eq:omega_prop_a}) and the following weak invariance notion:
\begin{equation}
\label{eq:omega_prop_aweaks}
\tilde{\rhmap}(\tilde{\bfw}) \cap \Omega(\tilde{\phi}) \equiv \tilde{\rhmap}(\tilde{\bfw})  \cap \left( \Omega^{\bar{\theta}}(\tilde{\phi}) \times \left \{ \bar{\theta} \right \} \right)  \neq  \varnothing \quad \forall \tilde{\bfw}=[\bfw,\theta]\in \Omega(\tilde{\phi}).
\end{equation}
with $\bar{\theta}=\bmod(\theta+1,T)$.

\textit{Condition (\ref{eq:omega_prop_b})}:\hspace{1ex} Since $\phi(t+1) \in \rhmap(\phi(t),\Theta(t)) = (\better \circ \shift)(\phi(t),\Theta(t)) $, it follows from Assumption~\ref{assum_rhmap}\eqref{assum_rhmap_c} and the invariance of $V$ in (\ref{eq:Vdef}) with respect to $\shift$, that $V(\phi(t))$ is nonincreasing. This implies that a finite limit $\underline{V}\in\mathbb{R}_+$ exists and the first equality in (\ref{eq:omega_prop_b}) holds, since $V$ takes always nonnegative values. To verify the second equality in (\ref{eq:omega_prop_b}), it is sufficient to consider that, given the supposed continuity of $V$, the following holds 
for all $\tilde{\bfw} = [ \bfw, \theta] \in \Omega(\phi)$:
\begin{equation}
V( \bfw ) = V\left(\lim_{n\rightarrow\infty}\phi(t_n) \right) = \lim_{n\rightarrow\infty} V(\phi(t_n)) =  \lim_{t\rightarrow\infty} V(\phi(t)) = \underline{V}. 
\end{equation}
\textit{Condition (\ref{eq:omega_prop_c})}:\hspace{1ex}
From (\ref{eq:omega_prop_aweaks}), given $\tilde{\bfw} \in\Omega(\tphi)$, there exists $\tilde{\bfy} \in\Omega(\tphi)$ such that $\tilde{\bfy} \in \trhmap(\bfw)$. 
Since both $\tilde{\bfy} = [\bfy, \bar{\theta}]$ and $\tilde{\bfw} = [\bfw , \theta]$ belong to $\Omega(\tphi)$, we have $V(\bfy)=V(\bfw)$ from~\eqref{eq:omega_prop_b}.
Recalling that $\tilde{\rhmap}(\tilde{\bfw}) =  \rhmap(\bfw , \theta) \times \left \{ \bar{\theta} \right \} $, the following equivalent condition is verified:
\begin{equation}
\exists \bfy \in \rhmap(\bfw,\theta) : V(\bfy) = V(\bfw).
\end{equation}
Moreover, since $\rhmap(\bfw,\theta) = \better(\shift(\bfw),\theta)$  and $V$ is invariant with respect to $\shift$, with $V(\shift(\bfw)) = V(\bfw)$, the following holds for $\bfw^+=\shift(\bfw)$:
\begin{equation}
\exists \bfy \in \better(\bfw^+,\theta) : V(\bfy) = V(\bfw^+).
\end{equation}
From (\ref{eq:rh_map_c}) and the second equation in (\ref{eq:rh_map_d}), it follows
$\better(\bfw^+,\theta) = \left \{ \bfw^+ \right \} = \left \{ \bfy \right\}$ and $\tilde{\rhmap} (\tilde{\bfw}) = \left \{ \tilde{\bfy} \right \} = \left \{ [\bfy,\bar{\theta}] \right \}  \subseteq \Omega(\tilde{\phi})$, thus concluding the proof.
\end{proof}
Lemma \ref{lem:omega_prop} shows that each solution $\tilde{\phi}(\cdot)=[\phi(\cdot),\Theta(\cdot)]$, with $\phi(t)$ denoting the planned strategy of the agents at time $t$ over the time interval $\{t,\dots,t+T-1\}$, converges asymptotically to its associated $\omega$-limit set $\Omega(\tilde{\phi})$. On such set, the Lyapunov function $V$ takes always the same finite value $\underline{V}$.
Moreover, if any point $\tilde{\bfw}\in  \Omega(\tilde{\phi})$ is reached, the system remains in $\Omega(\tilde{\phi})$ and only the rotation operator $\shift$ is applied. This means that, at time $t+1$, the last element $\phi_{\{T-1\}}(t+1)$ of the new planned strategy corresponds to the first element $\phi_{\{0\}}(t)$ at time $t$, i.e. the state remains periodic over time with period $T$.
It is now possible to provide the main results on the feasibility and convergence of the state $\phi$ and output $\psi$ of the dynamical system (\ref{eq:state})-(\ref{eq:output}) describing the proposed receding horizon strategy.

\begin{thm}
	\label{thm_convergwardrop}
	Consider $\tilde{\phi}(\cdot) =[\phi(\cdot),\Theta(\cdot)] \in \Phi(\bfx^0)$ and associated output $\psi$, as presented in Definition \ref{def_solset}, and let Assumption~\ref{assum_rhmap} hold. If $\bfx^0 \in \setKper_{\langle 0,0 \rangle}$, it holds:
	\begin{subequations}
	\begin{equation}
	\label{eq:thmconv_1}
	\phi(t) \in \setKper_{\langle t,0 \rangle} \qquad \forall t\in \mathbb{N}
	\end{equation}
	\begin{equation}
	\label{eq:thmconv_2}
	\lim_{t\rightarrow\infty} \min_{\bfx\in\setWardper_{\langle t,0 \rangle}}	\left\|\phi(t) - \bfx \right\| = 0. 
	\end{equation}
	\begin{equation}
	\label{eq:thmconv_3}
	\lim_{t\rightarrow\infty} \min_{\bfx\in\setWardper_{\langle t,0 \rangle}}	\left\|\teta_{\langle t,0 \rangle} -\bfx \right\| = 0. 
	\end{equation}
	\end{subequations}
\end{thm}
\begin{proof}
See \ref{sec_convergwardrop}.
\end{proof}

Theorem \ref{thm_convergwardrop} establishes fundamental properties of the dynamical system (\ref{eq:state}) when the Lyapunov function $V$ and the update mapping $\better$ fulfil Assumption \ref{assum_rhmap}. In particular the state $\phi(t)$, i.e. the predicted strategy at time $t$ over the time horizon $\{t,\dots,t+T-1\}$, is always feasible, in the sense that it can be infinitely replicated over time and generate a periodic strategy which is feasible for the infinite-horizon game of Definition \ref{def_gamestatic}. Moreover, from (\ref{eq:thmconv_2}), the predicted strategy converges to the set of periodic aggregative equilibria $\setWardper$. This means that, at the limit, the predicted strategy $\phi(t)$ corresponds to the restriction over a single time period of some $\tilde{w}\in \setWardper$, i.e. a periodic equilibrium for the infinite-horizon game of Definition \ref{def_gamestatic}. Most importantly, this property not only holds for the predicted strategy $\phi(t)$ but also for the actual strategy $\psi_{\langle t,0 \rangle}$ implemented by the players under the considered receding horizon framework. It is worth emphasizing that these results also ensure a practical adaptability of the proposed control scheme to changes in the environment. In fact, for any change occurring in the game at some time $t$ (new players enter the game, current players leave the game or change their parameters), the receding horizon control scheme will eventually converge to the periodic equilibrium associated to the game with the new characteristics, as long as a feasible periodic strategy is considered at the initial time $t$.

\section{Example}
\label{sec_exgame}
A possible application of the receding-horizon control scheme introduced in Section \ref{sec_rhgame} is now presented, discussing the design of a better-response mapping $\better$ and a Lyapunov functional $V$ that fulfil Assumption $\ref{assum_rhmap}$ and ensure feasibility and convergence to equilibrium of the proposed control strategy. In particular, the problem of data routing over a multi-hop network is considered. The communication network is modelled as a graph $G=(\setV,\setL)$, where $\setV = \{1,\ldots,M\}$ and  $\setL \subseteq \setV\times\setV$ denote the sets of nodes and links in the graph, respectively.  
The network users are modelled as a finite population $\setN=\left \{ 1, \dots, N \right \}$ of self-interested agents. Each agent $i\in \mathcal{N}$ needs to periodically transmit (with period $T$) a certain amount of data $D_i$  from some sender node $v^s_i \in \setV$ to a receiver node $v^r_i \in \setV$. This task must be performed over a limited time interval of length $\eta_i \in  \{0,\dots,T-1\}$, namely $\{kT+\tcapi_i,\dots,kT+\tcapi_i+\tend_i\}$ with $\tcapi_i \in \{0,\dots,T-1\}$ and $k\in \mathbb{N}$.
Denoted by $\setP$ the set of all paths in the network, we represent by $x_{i,p}(t)\in [0,\mu_i] \subseteq \mathbb{R}_+$ the amount of data transmitted by player $i\in \mathcal{N}$ on path $p\in \setP$ at time $t\in \mathbb{N}$, where $\mu_i$ represents the maximum transmission rate of agent $i$ on a single path. This notation can be extended, denoting by $x_i(t) \in [0,\mu_i]^{|\setP|} = \setX_i$ the overall strategy of agent $i$ at time $t$, i.e. the vector of transmitted data on each path $p\in \setP$ at time $t$. As we are considering an infinite-time horizon, the full strategy of agent $i\in \mathcal{N}$ corresponds to a function $x_i : \mathbb{N} \rightarrow \setX_i$ and its feasibility set can be defined as follows:
\begin{equation}
\label{eq:setfeas_ex}
\setfeasinf_i := \left\{ x_i : \mathbb{N} \rightarrow \setX_i : x_{i\langle \tcapi_i,k \rangle} \in\setXbar_i,\quad \forall k \in \mathbb{N} \right\}.
\end{equation}
Under the present modelling framework with periodic constraints,
the feasible strategy $x_i$ evaluated on each time period $\{kT+\tcapi_i,\dots,(k+1)T+\tcapi_i-1\}$ must belong to some feasibility set $\setXbar_i$, with the following expression:
\begin{equation}\label{eq_feassetT_ex}
	\setXbar_i \triangleq \left\{z_i : \{0,\dots,T-1\} \rightarrow  [0,\mu_i]^{|\setP|} :\; \sum_{\tau = 0}^{\tend_i}\sum_{p \in \setP_i} z_{i,p}(\tau)= D_i \right\}.
\end{equation}
In other words, in the availability time interval $\{0,\dots,\tend_i\}$ within each time period $\{0,T-1\}$, the total amount of data $z_{i,p}$ transmitted by player $i$ (at each time $\tau \in \{0,\dots,T-1\}$ and on each path $p\in\setP_i$ connecting sending node $v^s_i$ and receiving node $v^r_i$) must be equal to the total information $D_i$ to be transmitted.
As in previous sections, we denote by $\setX = \prod_{i\in\setN}\setX_i$ the global action space, and by $\setfeasinf = \prod_{i\in\setN}\setfeasinf_i$ the set of global feasible strategies $x$. It is now possible to define the strategy aggregation function $\Lambda : \setX \rightarrow \mathbb{R}^{|\setL|}$.  For a certain global strategy $x(t)$ at time $t$, the $l$-th component of $\Lambda(x(t))$ --- denoted as $\Lambda_l(x(t))$ --- corresponds to the resulting total traffic on link $l$:
\begin{equation}
\Lambda_l(x(t)) \triangleq \sum_{i \in \setN} \sum_{p \in \setP_i} L_{l,p} \cdot x_{i,p}(t),
\label{eq_Lambdaelle}
\end{equation}
where $L_{l,p}=1$ if the link $l$ is included in path $p$ (and 0 otherwise).
It is assumed that the price of transmitting over a link $l$ depends on the traffic $\Lambda_l$ through a monotonically increasing function $\Pi\colon \mathbb{R} \to \mathbb{R}$.
\begin{assum}
\label{assum:Lipschitz}
The transmission price function $\Pi$ is Lipschitz continuous, with Lispchitz constant $\kappa_{\Pi}$.
\end{assum}
The price $\pi_p$ for transmitting over a certain path $p$ can now be expressed as the following sum:
\begin{equation}
\label{eq:pi_p}
\pi_p(\Lambda(x(t))) = \sum_{l\in \setL} L_{l,p} \cdot  \Pi( \Lambda_l(x(t))  ).
\end{equation}
The cost $J^{\infty}_i$ sustained by agent $i$ when a certain strategy $x$ is applied can then be defined as the average cost over an infinite-time horizon:
\begin{equation}
\Jinf_i(x_i,x) = \limsup_{t\rightarrow\infty}\frac{1}{t}\sum_{\tau=0}^{t-1} \sum_{p\in\setP_i} \pi_p \left(\Lambda(x(\tau)) \right) \cdot x_{i,p}(\tau).
\label{eq_indcostnetw}
\end{equation}
\subsection{Receding-horizon implementation}
It is now assumed that the strategy $x$ of the game is determined through the receding horizon characterized by system (\ref{eq:state}) in Section~\ref{sec_rhgame}.
The bold term $\bfx(t) \in \setX^T\subset \mathbb{R}^{N\cdot T \cdot |\setP|}$ in (\ref{eq:state1}) corresponds to the predicted strategy (i.e. the scheduled transmission of the players) over the time interval $\{t,\dots,t+T-1\}$. Consistently with previous notation, the term $\bfx_{\left\{\tau\right\}}(t)$, with $\tau\in\{ 0,\dots,T-1\}$, designates the strategy planned at time $t$ for time $t+\tau$, denoting by  $\bfx_{i\left\{\tau\right\}p}(t)$ the component associated to player $i$ and path $p$. The scalar quantity $\theta(t)$ in (\ref{eq:state2}) represents the offset of the current prediction window with respect to the chosen period $T$. Finally, the system output $y(t)$ in (\ref{eq:output}) corresponds to the strategy actually implemented at time $t$ (i.e. the data transmitted by each player at time $t$). Consistently with the envisioned receding horizon framework, $y(t)$ corresponds to the first time component $\bfx_{\{0\}}(t)$ of the planned strategy $\bfx(t)$.
The associated local cost $J_i$ of the player $i\in \mathcal{N}$ can be expressed as follows:
\begin{equation}
J_i(\bfx_i(t),\bfx(t)) = \sum_{\tau=0}^{T-1} \sum_{p\in\setP_i} \pi_p \left(\Lambda(\bfx_{\{\tau\}}(t)) \right) \cdot \bfx_{i\{\tau\}p}(t).
\label{eq_indcostlocal}
\end{equation}

To fully characterize the mapping $\rhmap$ that updates the planned strategy $\bfx(t)$ in the present case, the following expression is provided for the \textit{update} mapping $\better : \setX^T \times \{ 0,\dots,T-1\} \setmap \setX^T$ introduced in Definition \ref{def_rhmap}:\footnote{With a slight abuse of notation, the composition is carried out only with respect to the $\bfx$ argument.}
\begin{equation}
\better(\bfx,\theta) \triangleq (\better_{N}\circ\better_{N-1}\circ \cdots \circ \better_1)^{(\gamma)}(\bfx,\tcap).
\label{eq_defbettermap}
\end{equation}
Each submapping $\better_{i}$ corresponds to a strategy update by the $i$-th agent which, on the basis of the current state, reschedules part of its data traffic to a cheaper path and/or time interval. The whole sequence of individual strategy updates $\better_i$ is iterated over all players $i = 1,\ldots,N$ a finite number of times $\gamma\geq 1$.
When the mapping $\better_{i}$ is applied, player $i$ can potentially divert a certain amount of data (initially transmitted on path $\pda$ at time $\ubar{\tau}$) and transmit them on some other path $\pa$ at some other time $\bar{\tau}$. To evaluate which potential swaps are advantageous for agent $i$ at a certain state $\bfx$, the following quantity is introduced:
\begin{equation}\label{eq_Sn}
\begin{array}{rl}
S_i(\bfx,\tcap) = & \displaystyle \argmax_{(\ta,\tda,\pa,\pda)} \left (
\pi_{\ubar{p}} \left( \Lambda(\bfx_{\{\ubar{\tau}\}} ) \right) -
\pi_{\bar{p}}  \left( \Lambda(\bfx_{\{\bar{\tau}\}}  ) \right)
\right)  \left ( \mu_i - \bfx_{i\{\bar{\tau}\}{\pa}} \right) \bfx_{i\{\ubar{\tau}\}{\pda}} \\
\text{s.t. } & (\bar{\tau},\ubar{\tau}) \in\setA_i(\tcap) \\ 
			       & \pa,\pda\in \setP_i
\end{array}
\end{equation}
Note that the last two factors in the right-hand side of (\ref{eq_Sn}) are always nonnegative. This means that $S_i(\bfx,\tcap)$ will contain (if possible) tuples $(\ta,\tda,\pa,\pda)$ such that: \textit{(i)} the associated data swap is cost-advantageous (positivity of first factor), \textit{(ii)} agent $i$ can trasmit more data on path $\pa$ at time $\bar{\tau}$ (positivity of second factor), \textit{(iii)} agent $i$ can transmit less data on path $\pda$ at time $\ubar{\tau}$ (positivity of third factor).
In order to ensure feasibility of the updated strategy, the possible swap times pairs are restricted to the set $\setA_i(\tcap)$, which only includes time instants in the same availability interval $\{kT + \tcapi_i,\dots,kT+\tcapi_i+\tend_i\}$.
Through simple algebraic operations, the following expression can be derived for $\setA_i(\tcap)$:
\begin{equation}\label{eq_availab}
\setA_i(\tcap) \triangleq \bigcup_{\delta T \in \{ -T,0,T \}} \setAb(\tcapi_i-\tcap + \delta T,\tcapi_i+\tend_i-\tcap + \delta T)
\end{equation}
where $\setAb(t_1,t_2)$ returns the Cartesian product of the intersection between the time interval $\{t_1,\dots,t_2\}$ and the period $\{ 0,\dots,T-1\}$:
\begin{equation}
\setAb(t_1,t_2) \triangleq \left( \{t_1,\dots,t_2\} \cap \{ 0,\dots,T-1\}  \right) \times \left( \{t_1,\dots,t_2\} \cap \{ 0,\dots,T-1\}  \right).
\label{eq_availabpre}
\end{equation}
Recalling~\eqref{eq_Sn} and $\bar{\tcap}=\bmod(\tcap+1,T)$, the mapping $\better_{i}(\bfx,\tcap)$ can be expressed as:
\begin{equation}\label{eq_defbettermap_i}
\better_{i}(\bfx,\tcap) := \bigcup_{s_i \in \setS_i(\bfx,\bar{\tcap})} F(\bfx,s_i),
\end{equation}
where $\bar{\tcap}$ accounts for the fact that the mapping $\better_i$ operates on a predicted strategy $\bfx$ that has already been shifted forward in time by $\shift$.
Each $F(\bfx,s_i)\in \setX^{T}$ corresponds to a potential revised strategy for the overall population. For a certain $s_i=(\ta,\tda,\pa,\pda)\in \setS_i(\bfx,\tcap)$, its component $F_{j\left \{ \tau \right \} p }$ associated to player $j\in \mathcal{N}$, time $\tau\in \{0,\dots,T-1\}$ and path $p \in \setP$ has the following expression:
\begin{equation}
\label{eq:mapping_eq1}
F_{j\{\tau\}p}(\bfx,s_i) := \left \{
\begin{array}{ccc}
\bfx_{j\{\tau\}p} + \Delta(\bfx,s_i) &  \text{ if } &  j=i,\, p=\bar{p},\, \tau=\bar{\tau} \\
\bfx_{j\{\tau\}p} - \Delta(\bfx,s_i) &  \text{ if } &  j=i,\, p=\ubar{p},\, \tau=\ubar{\tau} \\
\bfx_{j\{\tau\}p} & & \text{otherwise}.
\end{array}
\right.
\end{equation}
In other words, when some $F(\bfx,s_i)\in \better_{i}(\bfx,\tcap)$ is selected, the player $i$ swaps an amount $\Delta(\bfx,s_i)$ of transmitted information from path $\pda$ at time $\tda$ to path $\pa$ at time $\ta$. The transmission schedule of all the other players remain unaltered. The quantity $\Delta(\bfx,s_i)$ has the following expression:
\begin{equation}
\Delta(\bfx,s_i) \triangleq \min\left\{ \left \lfloor \frac{\pi_{\pda} (\Lambda( \bfx_{\tda}) ) - \pi_{\pa} ( \Lambda(\bfx_{\ta}) )}{ \kappa_{\Pi}  (|\pa|+|\pda|)} \right \rfloor_+,\; \mu_{i}- \bfx_{\ta\{i\}\pa} ,\;
\bfx_{\tda\{i\}\pda} \right\}.
\label{eq_deltamax}
\end{equation}
where $\lfloor x \rfloor_+$ denotes positive part of $x$, $|p|$ quantifies the number of branches on path $p$ and $\kappa_{\Pi}$ is the Lipschitz constant of the transmission price function, as established in Assumption \ref{assum:Lipschitz}.
The first term in the $\min$ function in (\ref{eq_deltamax}) preserves the cost order between the two pairs of path and time instant considered for the swap whereas the other two terms ensure feasibility of the updated transmission schedule.

We wish to emphasize that the proposed mapping $\better$ has been designed in order to facilitate its distributed implementation in realistic contexts. In fact, the application of the submapping $\better_i$ can correspond to the independent action of agent $i$ that, in response to some broadcast of the transmission prices, decides to modify its predicted strategy in order to reduce its cost. This procedure can then be iterated $\gamma$ times over the whole population, thus obtaining the strategy update described by $\better$ in (\ref{eq_defbettermap}).

\subsection{Feasibility and convergence results}
It is now shown that the theoretical results presented in the previous section are valid for the present data transmission examples. To this end, the following Lyapunov function is considered:
\begin{equation}
\label{eq:Lyap_fun}
V(\bfx)= \sum_{\tau=0}^{T-1} g  (\bfx_{\{ \tau \}})  = \sum_{\tau=0}^{T-1} \sum_{l\in \setL} \Gamma(\Lambda_l(\bfx_{\{ \tau \}})) 
\end{equation}
with $\Gamma(x)$ defined as:
\begin{equation}
\label{eq:Gamma}
\Gamma(x) \triangleq \int_0^{x} \Pi(s) \, ds.
\end{equation}
Note that $\Gamma$ corresponds to the potential function associated to the price $\Pi$ on a single branch. As a result, $V$ can be interpreted as the \emph{global potential function} of the game \citep{NisanEtAl_AlgGTbook}.

\begin{prop}
The mapping $\better$ and the Lyapunov function $V$ introduced in (\ref{eq_defbettermap}) and (\ref{eq:Lyap_fun}), respectively, fulfil Assumption \ref{assum_rhmap}.
\label{prop_rhmap}
\end{prop}
\begin{proof}
See \ref{sec:rhmap}.
\end{proof}
\begin{prop}
The strategy $\psi(\cdot)$ actually implemented under the proposed receding horizon approach, with $\psi(t)=\phi_{ \{0 \} }(t)$ for all $t\in \mathbb{N}$, is a feasible infinite-horizon strategy. Equivalently, recalling the feasibility set $\setfeasinf = \prod_{i\in\setN}\setfeasinf_i$ with $\setfeasinf_i$ defined in (\ref{eq:setfeas_ex}), it holds $\psi(\cdot) \in \setfeasinf$.
\end{prop}
\begin{proof}
Let us denote as $\xi(t)$ the concatenation of the implemented strategy $\psi$ up to time $t-1$ with the predicted strategy $\phi$ at time $t$. The following expression can be provided for $\xi$:
\begin{equation}
\label{eq:xi_expr}
\xi_{\{\tau\}}(t) = \left \{
\begin{array}{ccc}
\psi(\tau) & \text{if } & \tau \in \{0,\dots,t-1\} \\
\phi_{\{\tau-t\}}(t) & \text{if } & \tau \in \{t,\dots,t+T-1\} \\
\end{array}
\right .
\end{equation}
Recalling expressions (\ref{eq:setfeas_ex}) and (\ref{eq_feassetT_ex}) and considering compactness of the constraints, the proposition statement is verified if the following holds:
\begin{equation}
\label{eq:induct}
\sum_{\tau=kT+\tcapi_i}^{kT+\tcapi_i+\eta_i}\sum_{p\in\setP_i} \xi_{i\{\tau\}p}(t) = D_i, \qquad  \forall t\in \mathbb{N}, \, k: kT+\tcapi_i+\eta_i \leq t+T-1,\,\forall i\in\setN
\end{equation}
This can be verified by induction over $t$, considering that (\ref{eq:induct}) holds by definition at time $t=0$ since $\xi(0)=\phi(0) \in \setKper_{\langle 0,0 \rangle} \subseteq \setfeasinf_{\langle 0,0 \rangle}$. It is now assumed that $\xi(t-1)$ fulfils (\ref{eq:induct}) and the same is proved for $\xi(t)$. Considering $\theta=\Theta(t-1)$ and $\bar{\theta}=\Theta(t)$ as the phase associated to the considered times $t-1$ and $t$, from expression (\ref{eq:xi_expr}) it is sufficient to verify the following:
\begin{equation}
\sum_{\tau=0}^{\tcapi_i+\eta_i-\bar{\theta}} \sum_{p\in\setP_i} \phi_{i\{\tau\}p}(t) = \sum_{\tau=0}^{\tcapi_i+\eta_i-\bar{\theta}} \sum_{p\in\setP_i} \phi_{i\{\tau+1\}p}(t-1) \qquad \forall i \in \mathcal{N} : \tcapi_i\leq \bar{\theta} \leq \tcapi_i+\eta_i
\end{equation}
Recalling that $\phi(t)=\rhmap(\phi(t-1),\Theta(t-1)) = \better(\shift(\phi(t-1)),\Theta(t-1))$ and considering $\bfx=\shift(\phi(t-1))$, from (\ref{eq_defbettermap}) it is sufficient to show the following:
\begin{equation}
\label{eq:final_cond}
\begin{array}{cc}
\displaystyle
\sum_{\tau=0}^{\tcapi_i+\eta_i-\bar{\theta}} \sum_{p\in\setP_i} \bfy_{i\{\tau\}p} = \sum_{\tau=0}^{\tcapi_i+\eta_i-\bar{\theta}} \sum_{p\in\setP_i} \bfx_{i\{\tau\}p} &
\begin{array}{c}
 \forall i \in \mathcal{N} : \tcapi_i\leq \bar{\theta} \leq \tcapi_i+\eta_i \\
 \forall \bfy \in \better_i(\bfx,\theta)
\end{array}
\end{array}
\end{equation}
From (\ref{eq:mapping_eq1}) and the expression of $\setA(\bar{\theta})$ in (\ref{eq_availab}), all transmission swaps associated to $\better_i$ correspond to zero-sum variations in the considered time interval. It follows that (\ref{eq:final_cond}) and (\ref{eq:induct}) hold, thus concluding the proof.
\end{proof}

From the above results we can conclude that the \emph{update mapping} $\better$ proposed in (\ref{eq_defbettermap}) --- corresponding to each player iteratively updating its predicted strategy in order to reduce its local transmission cost --- is associated to a reduction of the \emph{global potential function} $V$ in (\ref{eq:Lyap_fun}). It follows that all the properties established in Theorem \ref{thm_convergwardrop} are also valid in the present case. In particular, the predicted strategy $\phi(t)$ and the actual implemented strategy $\psi_{\langle t,0 \rangle}$ (introduced in Definition \ref{def_solset}) are feasible and asymptotically converge to a periodic (aggregative) equilibrium, thus representing a stable outcome of the data transmission game. Note that such a result also ensures an intrinsic adaptability of the proposed control strategy. Thanks to the envisioned receding horizon approach, if a change occurs in the game over time (players enter/exit/change their task parameters) the proposed update strategy will ensure seamless convergence to the new associated equilibrium.

\subsection{Simulation results}
\begin{figure}
	\centering
	\def\svgwidth{0.6\columnwidth}
	\large{
		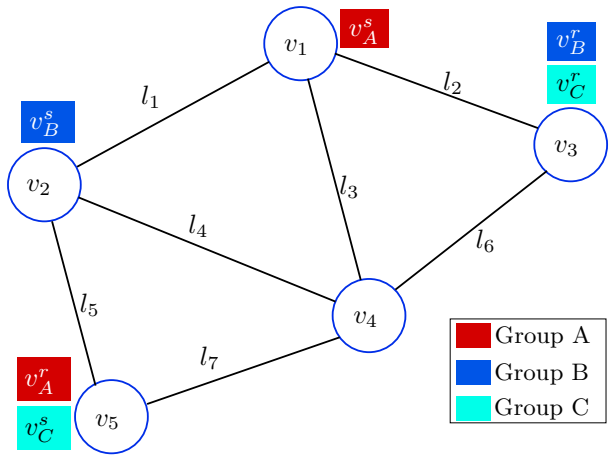
	}
	\caption{Topology of the considered network. The population of users is uniformly distributed into three different source-destination pairs.}
	\label{fig_netw}
\end{figure}

The performance of the proposed receding-horizon control scheme is evaluated on the network
represented in Figure~\ref{fig_netw}. The considered population $\mathcal{N}$ of $N=100$ transmitting agents is randomly repartitioned in three groups, with $\mathcal{N} = \mathcal{N}_A \cup  \mathcal{N}_B \cup  \mathcal{N}_C$. For simplicity, it is assumed that all users in each group are associated to the same pair of sending and receiving node, as represented in Figure~\ref{fig_netw}.

Recalling the constraints formulation in (\ref{eq_feassetT_ex}), each agent $i$ needs to periodically transmit a total amount of data $D_i$ in the interval $\{ kT+\tcapi_i, \dots, kT+\tcapi_i+\eta_i\}$, for all $k\in \mathbb{N}$. In the present case study, the constraints period has been chosen as $T=7s$ and the maximum transmission rate $\mu_i$ is supposed to be equal for all players, with $\setX_i = [0, \mu_i]=[0,1]$ for all $i\in \mathcal{N}$. The values of $D_i$, $\eta_i$ and $\tcapi_i$ have been extracted from uniform distributions whose support ensures non-emptiness of the feasibility set $\setXbar_i$. The transmission price $\Pi$ over a single link is assumed to be linear, with $\Pi(d)= k d$. Moreover, in addition to the data transmitted by the coordinated agents, further information is transmitted by external sources on some network paths.

At time $t=0$ the agents start with some initial feasible predicted strategy $x^0$ over the time interval $\{0,\dots,T-1\}$. The receding horizon strategy update presented in previous sections is then applied: at each time instant $t$ the agents create a new planned strategy $\shift(\phi(t-1))$ by simply shifting by one time step their planned strategy $\phi(t-1)$ at time $t-1$. The mapping $\better$ is then applied, with the individual agents iteratively updating their transmission schedule in order to reduce their cost, thus obtaining the revised strategy $\phi(t)=\better(\shift(\phi(t-1),\Theta(t-1))$ at time $t$ over the time horizon $\{t,\dots,t+T-1\}$. The values in the first step are actually implemented, with $\psi(t)=\phi_{\{ 0\}}(t)$ and the whole procedure is repeated in a receding horizon fashion. A system fault is considered at time $t=85\text{ s}$, with all the agents $\mathcal{N}_B$ (representing 25\% of the overall population in this example) becoming isolated from the system and unable to transmit until the fault is resolved at time $t=125 \text{ s}$.

\begin{figure}[tb]
	\centering
	\includegraphics[trim={0cm 0 0.5cm 0},clip,width=0.8\columnwidth]{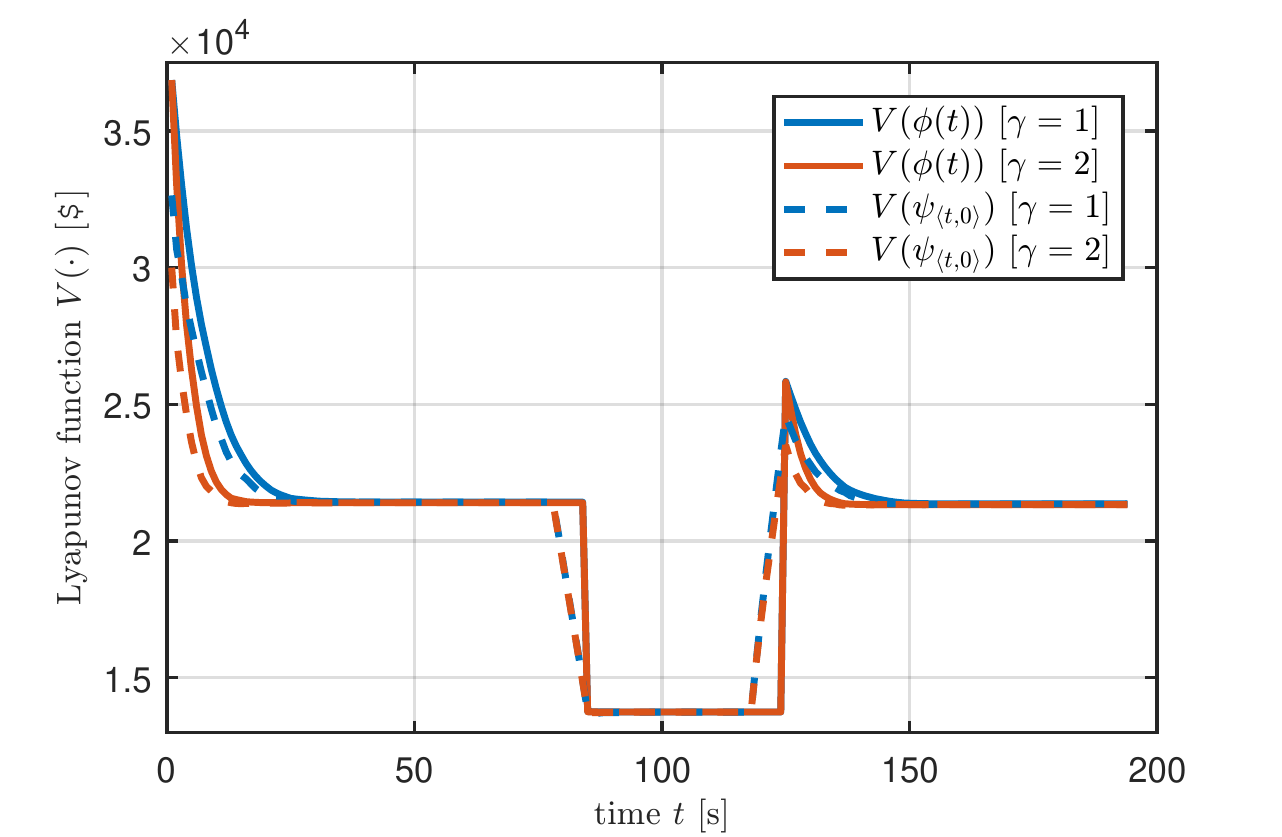}
	\caption{Values of the Lyapunov function across time evaluated over the predicted strategy (solid lines) and implemented strategy (dashed lines).} 
	\label{fig_Lyap}
\end{figure}

\begin{figure}[tb]
	\centering
	\includegraphics[trim={1.8cm 1.5cm 1cm 0},clip,width=\columnwidth]{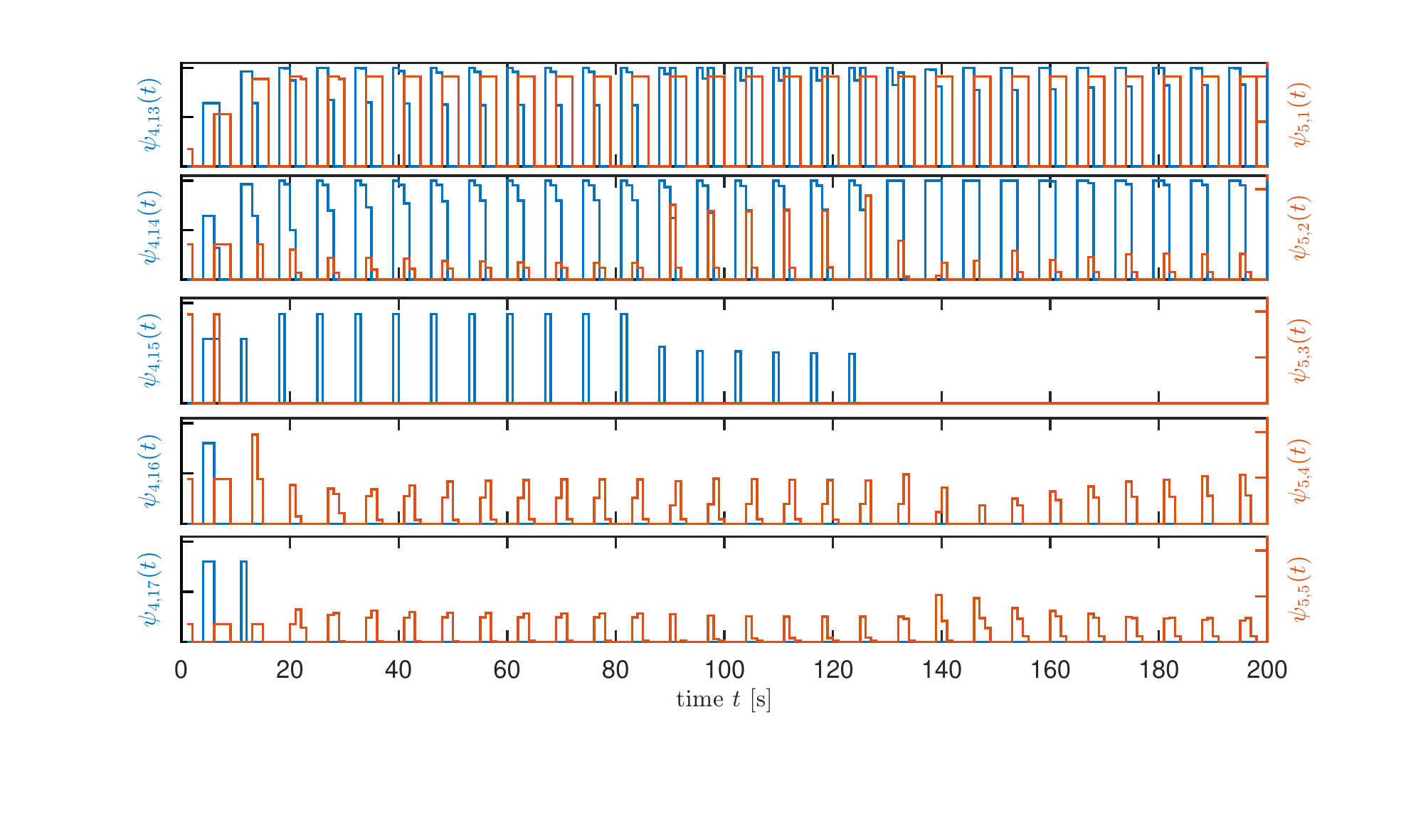}
	\caption{Implemented transmission rate $\psi_{i,p}(t)$ by players $i=4$ (Type C) and $i=5$ (Type A) over relevant paths $p$.}%
	\label{fig_strat}
\end{figure}

\begin{figure}[tb]
	\centering
	\includegraphics[trim={0 0 0 0},clip,width=0.8\columnwidth]{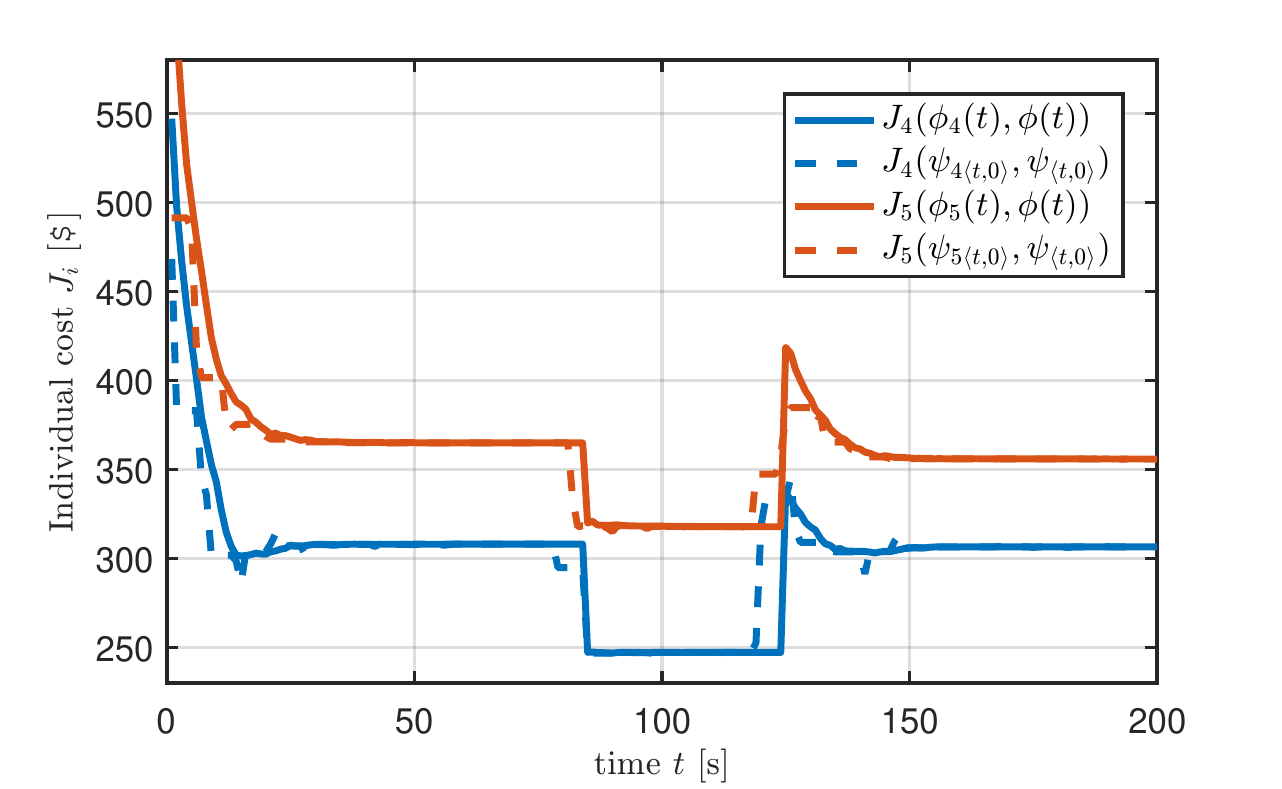}
	\caption{Cost $J_i$ incurred by the individual players $i=4$ (type C) and $i=5$ (type A), evaluated over their predicted (continuous lines) and implemented strategies (dashed lines).}
	\label{fig_indcost}
\end{figure}

\begin{figure}[tb]
	\centering
	\includegraphics[trim={0 0 0 0},clip,width=0.8\columnwidth]{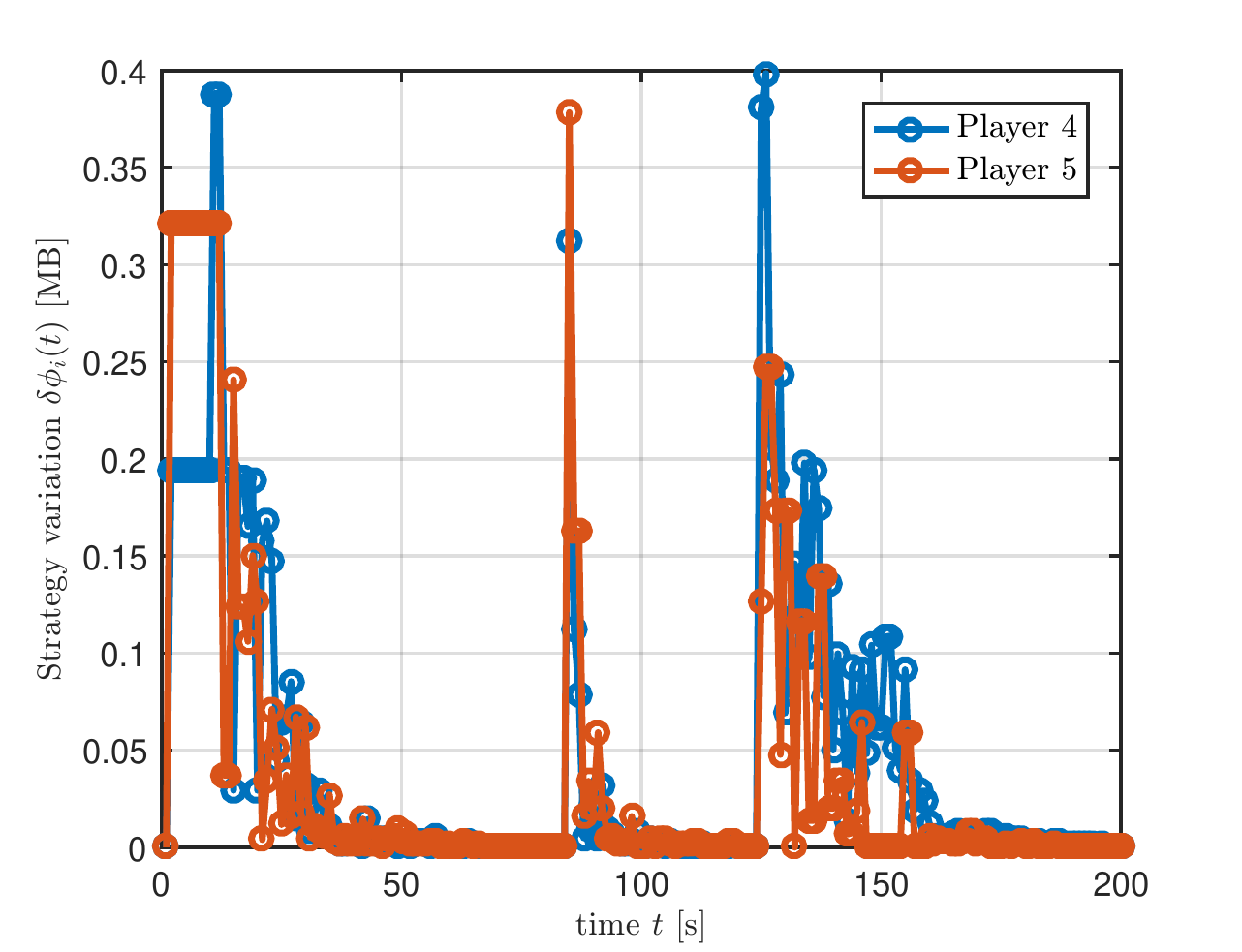}
	\caption{Strategy variation $\delta\phi(t)_i$ for players $i=4$ (type C) and $i=5$ (type A).}
	\label{fig_deltas}
\end{figure}

The discussed scenario has been simulated over the time interval $[0\text{ s},200\text{ s}]$, verifying the results of the proposed control strategy in the pre-fault interval $[0\text{ s},85\text{ s}]$, during the fault $[85\text{ s},125\text{ s}]$ and after the fault repair $[125\text{ s},200\text{ s}]$. Two different values of $\gamma$ in~\eqref{eq_defbettermap} are considered, namely for the cases that one and two better-response updates are iteratively performed by the whole population. The Lyapunov function (defined in~\eqref{eq:Lyap_fun} as the \mbox{\emph{global potential function} of} the system) has been evaluated over the predicted strategy $\phi(t)$ and over the actual implemented strategy $\psi(t)$. The corresponding values $V(\phi(t))$ and $V(\psi_{\langle t,0 \rangle})$ are shown in Figure~\ref{fig_Lyap}. It can be seen that a minimum value is reached in all three cases (pre/during/past fault), implying convergence to an infinite-horizon periodic equilibrium. As expected, convergence is faster when $\gamma=2$ as a higher number of strategy updates is considered on the single prediction horizon.
The implemented strategy $\psi$, i.e. the transmission rate of each agent on each network path, is represented in Fig. \ref{fig_strat} for two sample players $i=4$ (type C) and $i=5$ (type A), with $\gamma=1$. Consistently with previous considerations, it can be seen that a periodic transmission schedule (corresponding to a periodic equilibrium) is obtained in all the considered time intervals (pre/during/past fault). It is worth emphasizing that, for $t=[0\text{ s},85\text{ s}]$ and $t=[125\text{ s},200\text{ s}]$, two different equilibria are reached (even if the number and parameters of players in the two intervals are equal).
The transmission cost of the devices on a single period, defined in (\ref{eq_indcostlocal}), has been evaluated over the planned and implemented strategy of players $i=4$ and $i=5$. The corresponding values of $J_i(\phi_i(t),\phi(t))$ and $J_i(\psi_{i\langle t, 0 \rangle},\psi_{\langle t, 0 \rangle})$ are shown in Figure \ref{fig_indcost}. In addition to the significant changes following the fault at $t=85 \text{ s}$ and the repair at $t=125\text{ s}$, it can be seen that the transmission costs are lower when the players of type B are not operating and the paths are less congested.
Finally, the strategy variations of the players across time, expressed as the amount of rescheduled transmission data $\delta\phi_i(t)=||\phi_{i}(t) - \phi_{i}(t-1) ||$ at each time instant $t$, are shown in Figure \ref{fig_deltas}. Substantial peaks can be seen at fault time $t=85\text{ s}$ and at repair time $t=125\text{ s}$, as the player react to a significant modification of the game parameters. As expected, the norm of strategy updates tends to zero as the periodic equilibrium is reached in the three time intervals.

\section{Conclusions}
\label{sec:conc}
The paper proposes a novel framework for aggregative games, considering players that minimize their cost functions over an infinite-time horizon, subject to heterogeneous periodic constraints. The associated aggregative equilibrium concept is introduced and a receding horizon mechanism is proposed for the coordination of the agents. This novel setup allows to consider a wide array of realistic scenarios, including cases where multiple agents need to periodically perform a certain task over potentially overlapping time windows.
Sufficient conditions for feasibility and convergence to equilibrium of receding-horizon control schemes are derived. This is obtained by introducing a discrete-time multi-valued dynamical system describing the time evolution of the strategies that are planned and actually implemented by the players, and by using proper invariance-theory tools to study the limit behaviour of the associated solutions. These results are independent of the initial planned strategy: following a change in the number and/or parameters of players, convergence to the new associated equilibrium is still ensured. The example of data routing over a multi-hop network is used to demonstrate the design of distributed better-response schemes that can be applied in the considered receding horizon context and to evaluate the performance of the proposed approach.

Future work will analytically characterize the robustness and disturbance rejection properties of the proposed scheme with respect to persistent noise, externalities and players behaviour. The capability of the scheme of estimating and predicting global system quantities that affect the players' objective functions will also be investigated. Finally, more general characterizations of the players' cost functions, extensions to non-periodic frameworks and the application of alternative distributed control schemes will also be evaluated.

\appendix
\section{Proof of Theorem \ref{thm_convergwardrop}.}
\label{sec_convergwardrop}
	\textit{Condition (\ref{eq:thmconv_1})}: this is proved by induction. Since the condition is verified for $t=0$ by hypothesis, it is sufficient to show that the following holds when $\phi(t) \in \setKper_{\langle t,0 \rangle}$:
\begin{equation}
\label{eq:thmconv_pr1}
\rhmap(\phi(t),\Theta(t)) = (\better \circ \shift)(\phi(t),\Theta(t)) = \better (\shift(\phi(t)),\Theta(t)  )  \subseteq \setKper_{\langle t+1,0 \rangle}.
\end{equation}
Considering Assumption~\ref{assum_rhmap}\eqref{assum_rhmap_b} and the fact that $\setKper_{\langle t+1,0 \rangle} = \setKper_{\langle \Theta(t+1),0 \rangle}$ from (\ref{eq:Ktilde}), condition (\ref{eq:thmconv_pr1}) holds if $\shift(\phi(t)) \in \setKper_{\langle t+1,0 \rangle}$. Since $\phi(t) \in \setKper_{\langle t,0 \rangle}$, there exists $ \tilde{z} \in \setKper$, such that $\phi(t) = \tilde{z}_{\langle t,0 \rangle} \in \setKper_{\langle t,0 \rangle} $. By construction, we have $\shift(\phi(t))= \tilde{z}_{\langle t+1,0 \rangle} \in \setKper_{\langle t+1,0 \rangle}$, which concludes the proof.

\textit{Condition (\ref{eq:thmconv_2})}: The following equivalent set of conditions is analysed:
\begin{subequations}
	\label{eq:thmconv_2alt}
\begin{equation}
\label{eq:thmconv_2alta}
\lim_{t\rightarrow\infty} \min_{\bfx\in \Omega^{\Theta(t)}(\tphi)}	\left\|\phi(t) - \bfx \right\| = 0. 
\end{equation}
\begin{equation}
\label{eq:thmconv_2altb}
\Omega^{\Theta(t)}(\tphi) \subseteq \setWardper_{\langle t,0 \rangle} .
\end{equation}
\end{subequations}
where $\Omega^{\Theta(t)}(\tphi)$ corresponds to (\ref{eq_omegalimset_tau}) evaluated at $\theta=\Theta(t)$.
To check that (\ref{eq:thmconv_2alta}) holds, note that the verified condition (\ref{eq:omega_prop_a}) can equivalently be written as:
\begin{equation}
\label{eq:limit_expr}
\lim_{t\rightarrow\infty} \min_{\tilde{\bfx}=[\bfx,\theta]\in \Omega(\tphi)}	\left\|\tilde{\phi}(t) - \tilde{\bfx} \right\| =  \left \|
\left [
\begin{array}{c}
\phi(t) \\ \Theta(t)
\end{array}
\right]  -  \left [
\begin{array}{c}
\bfx \\ \theta
\end{array}
\right] \right \| = 0.
\end{equation}
Since $\theta$ and $\Theta(t)$ always take integer values in $\{0,\dots,T-1\}$, from (\ref{eq:limit_expr}) there exists a sufficiently large $\bar{t}$ such that:
\begin{equation}
\label{eq:limit_expr2}
\resizebox{0.99\hsize}{!}{$
	\displaystyle
\min_{\tilde{\bfx}=[\bfx,\theta]\in \Omega(\tphi)}	 \left \|
\left [
\begin{array}{c}
\phi(t) \\ \Theta(t)
\end{array}
\right]  -  \left [
\begin{array}{c}
\bfx \\ \theta
\end{array}
\right] \right \| = \min_{\bfx \in \Omega^{\Theta(t)}(\tphi)} \left \|
\left [
\begin{array}{c}
\phi(t) \\ \Theta(t)
\end{array}
\right]  -  \left [
\begin{array}{c}
\bfx \\ \Theta(t)
\end{array}
\right] \right \| \quad \forall t\geq \bar{t}
$}
\end{equation}
As the right-hand side in (\ref{eq:limit_expr2}) tends to zero when $t \rightarrow +\infty$ from (\ref{eq:limit_expr}), it follows that (\ref{eq:thmconv_2alta}) holds. To check (\ref{eq:thmconv_2altb}), first note that $\Omega^{\Theta(t)}(\tphi)\subseteq \setKper_{\langle t,0 \rangle}$ as a result of (\ref{eq:omega_prop_a}), (\ref{eq:thmconv_1}) and compactness of the considered constraints. Therefore, any $\bfw \in \Omega^{\Theta(t)}(\tphi)$ can be associated to a periodic strategy $\tilde{w} : \mathbb{N} \rightarrow \setX^T$ with the following properties:
\begin{equation}
\label{eq:Omega_incl}
\tilde{w}_{\langle \Theta(t) + \tau ,0 \rangle}=\rhmap^{(\tau)}(\bfw) = \left \{ \shift^{(\tau)} (\bfw)\right \} \subseteq \Omega^{(\Theta(t+\tau))} (\tphi) \quad \forall \tau \in \mathbb{Z}
\end{equation}
where the superscript $(\tau)$ denotes $\tau$ iterated applications of the mapping $\rhmap$ or $\shift$ and the set inclusion straightly follows from (\ref{eq:omega_prop_c}). From Definition \ref{def_rhmap}, it holds:
\begin{equation}
\better \left ( \tilde{w}_{\langle \theta ,0 \rangle} \right) = \left \{ \tilde{w}_{\langle \theta,0 \rangle} \right \} \qquad \forall \theta \in \{0,\dots,T-1\}
\end{equation}
Considering the first equation in (\ref{eq:rh_map_d}) yields:
\begin{equation}
J_i(\tilde{w}_{i\langle \tcapi_i,0 \rangle},\tilde{w}_{\langle \tcapi_i,0 \rangle}) = \min_{y_i \in \setXbar_i} J_i(y_i,\tilde{w}_{i\langle \tcapi_i,0 \rangle}) \qquad \forall i \in \mathcal{N}, \, \, \forall \tcapi_i \in \{0,\dots,T-1\}
\end{equation}
which from (\ref{eq_setfeasinf}) and (\ref{eq:Ktilde}) can equivalently be written as:
\begin{equation}
\label{eq:Jrel}
J_i(\tilde{w}_{i\langle 0,k \rangle},\tilde{w}_{\langle 0,k \rangle}) = \min_{y_i \in \setKper_{i\langle 0,k\rangle}} J_i(y_i,\tilde{w}_{i\langle 0,k \rangle}) \qquad \forall i \in \mathcal{N}
\end{equation}
Note that $\tilde{w}\in\setWardper$ from definition (\ref{eq:Wtilde}), thus concluding the proof.

\textit{Condition (\ref{eq:thmconv_3})}: Having verified (\ref{eq:thmconv_2}), it is sufficient to prove the following:
\begin{equation}
\label{eq:state_output_limit}
\lim_{t\rightarrow\infty} \left\|\psi_{\langle t,0 \rangle} - \phi(t) \right\| = 0.
\end{equation}
In this respect, consider any $\tilde{\bfw}=[\bfw,\theta] \in\Omega(\phi)$.
Following~\eqref{eq_omegalimset}, let $\setF(\phi,\bfw)\triangleq\{\{t_n\}_{n\in\mathbb{N}}:\, \lim_{n\to\infty}t_n = \infty,\, \lim_{n\to\infty}\phi(t_n) = \bfw\}$ be the nonempty set of all time sequences over which the solution $\phi$ converges to $\bfw$.
Then, from the established upper semicontinuity of $\rhmap$ we have~\cite[Chap.~5]{RockafellarWets98}
\begin{multline*} 
\bigcup_{\{t_n\}\in\setF(\phi,\bfw)}\limsup_{n\rightarrow \infty} \rhmap(\phi(t_n),\Theta(t_n) ) \\
= \left\{\bfy:\, \exists \phi(t_n)\rightarrow \bfw,\, \exists \bfy_n \rightarrow \bfy,\, \bfy_n \in\rhmap(\phi(t_n),\Theta(t_n)) \right\}
\subseteq \rhmap(\bfw,\theta) .
\end{multline*}
From (\ref{eq:omega_prop_c}) and $[\bfw,\theta] \in \Omega(\phi)$ we have $\rhmap(\bfw,\theta) = \{\shift(\omega)\}$. Since the mapping $\rhmap$ is single-valued in $\bfw$, we can write:
\begin{equation*}
\bigcup_{\{t_n\}\in\setF(\phi,\omega)}\limsup_{n\rightarrow \infty} \rhmap(\phi(t_n),\Theta(t_n))
= \lim_{n\rightarrow \infty} \rhmap(\phi(t_n),\Theta(t_n)) = \rhmap(\bfw,\theta)
\end{equation*}
which corresponds to continuity of $\rhmap(\bfw,\theta)$ on  $[\bfw,\theta]  \in\Omega(\tphi)$~\citep{RockafellarWets98}. As a result, it holds:
\begin{equation*}
\lim_{n\rightarrow\infty} \phi(t_n+1) \overset{(a)}{=} \lim_{n\rightarrow\infty} \rhmap(\phi(t_n),\Theta(t_n)) = \rhmap(\bfw,\theta) \overset{(b)}{=}
\{\shift(\omega)\}
\end{equation*}
where equality $(a)$ follows from~\eqref{eq_solset}, and $(b)$ from (\ref{eq:omega_prop_c}). Hence, recalling $\teta(t) = \phi_{\{0\}}(t)$ for all $t\in\mathbb{N}$, we have
\begin{align*}
\lim_{n\rightarrow\infty} & \teta(t_n) = \bfw_{\{0\}},\\
\lim_{n\rightarrow\infty} & \teta(t_n+1)  = \shift_{\{0\}}(\bfw) = \bfw_{\{1\}},\\
& \qquad \vdots \\
\lim_{n\rightarrow\infty} & \teta(t_n+T-1)  = \shift^{(T-1)}_{\{0\}}(\bfw)= \bfw_{\{T-1\}},
\end{align*}
and then
\begin{equation}
\label{eq:limit}
\lim_{n\rightarrow\infty} [\teta(t_n), \teta(t_n+1), \cdots, \teta(t_n+T-1)] = \lim_{n\rightarrow\infty} \teta_{\langle t_n,0 \rangle} = \bfw. 
\end{equation}
Since (\ref{eq:limit}) holds for any $[\bfw,\theta] \in \Omega(\phi)$ and any $t_n\rightarrow +\infty$ such that $\phi(t_n) \rightarrow \bfw$, it follows that (\ref{eq:state_output_limit}) and (\ref{eq:thmconv_3}) are verified, thus concluding the proof.

\section{Proof of Proposition \ref{prop_rhmap}.}
\label{sec:rhmap}
\textit{Assumption \ref{assum_rhmap}-\eqref{assum_rhmap_a}}:\hspace{1ex} It is first shown that each submapping $\better_i$ in (\ref{eq_defbettermap_i}) is upper semi-continuous. In this respect, it is sufficient to prove that the graph $\mathbf{grf}(\better_i)$ of $\better_i$ is closed~\citep{Aliprantis06}, where $\mathbf{grf}(\better_i)$ has the following expression:
\begin{multline*}
\mathbf{grf}(\better_i) \triangleq \left\{
(\bfx,\tcap,\bfy):\, \bfx\in\setX^T,\, \theta \in \{0,\dots,T-1\} ,\, \bfy\in\better_i(\bfx,\tcap)\right\} \\
= \bigcup_{\tcap\in\{0,\dots,T-1\}}\bigcup_{s_i\in(\setA_i(\bar{\theta})\times\setP_i \times\setP_i)}\left\{\left(\bfx,\theta,\bfy\right):\,
\bfx\in \Xi(\bar{\theta},s_i),\, \bfy \in F(\bfx,s_i) \right\} ,
\end{multline*}
with $\bar{\theta}=\bmod(\theta+1,T)$ and
$\Xi(\bar{\theta},s_i) \triangleq \{\bfx\in\setX^T:\, s_i\in\setS_i(\bfx,\bar{\tcap})\}$.
Given $s=(\ta,\tda,\pa,\pda)$, the function maximized in (\ref{eq_Sn}) is denoted as $\lambda(\bfx,s)$:
\begin{equation}
\label{eq:lambda}
\lambda(\bfx,s)=\left (
\pi_{\ubar{p}} \left( \Lambda(\bfx_{\{\ubar{\tau}\}} ) \right) -
\pi_{\bar{p}}  \left( \Lambda(\bfx_{\{\bar{\tau}\}}  ) \right)
\right)  \left ( \bar{\mu}_i - \bfx_{i\{\bar{\tau}\}{\pa}} \right) \bfx_{i\{\ubar{\tau}\}{\pda}}
\end{equation}
The following equivalent expression can then be provided for $\Xi$:
\begin{equation}
	\Xi(\bar{\theta},s_i) = \{\bfx\in\setX^T:\, \lambda(\bfx,s_i) \geq\lambda(\bfx,s),\, \forall s\in \setA_i(\bar{\theta})\times\setP_i \times\setP_i \}.
\end{equation}
One can verify that the set $\Xi(\bar{\theta},s_i)$ is always closed (defined by a set of non-strict inequalities) and $F(\bfx,s_i)$ is continuous with respect to $\bfx$ for any $s_i$ (from expression \eqref{eq:mapping_eq1} of its individual component). It follows that $\mathbf{grf}(\better_i)$ is closed (union of finite number of closed sets), implying upper semi-continuity of $\better_i$ and of $\better$ (composition of a finite number of upper semi-continuous mappings $\better_i$).

\textit{Assumption \ref{assum_rhmap}-\eqref{assum_rhmap_b}}:\hspace{1ex} Consider $\bar{\theta}=\bmod(\theta+1,T)$. From (\ref{eq:Ktilde}) and (\ref{eq_feassetT_ex}), the condition $\bfx\in\setKper_{\langle \bar{\theta},k \rangle}$ in the present case can equivalently be expressed as:
\begin{subequations}
	\label{eq:cond_x}
	\begin{equation}
	\label{eq:max_x}
	\bfx_i \in [0,\mu_i]^{T|\setP|} \qquad \forall i\in \mathcal{N}.
	\end{equation}
	\begin{equation}
	\label{eq:totsum_x}
	\sum_{\tau\in{\mathcal{T}}} \sum_{p\in \setP_i} \bfx_{i\{\tau\} p}  = D_i \qquad \forall i\in \mathcal{N}.
	\end{equation}
\end{subequations}
where $\mathcal{T}$ represents the intersection (in the coordinates of the current prediction horizon) between the periodic availability interval $\{kT+\tcapi_i,\dots,kT+\tcapi_i+\eta_i\}$ and the current planning horizon window $\{\bar{\theta},\dots,\bar{\theta}+T-1\}$:
\begin{equation}
\mathcal{T} = \bigcup_{\delta T \in \{ -T,0,T \}} \{\tcapi_i-\bar{\theta}+\delta t , \dots, \tcapi_i-\bar{\theta}+\eta_i+\delta t\} \cap \{0,\dots T-1 \}  
\end{equation}
Similarly, to check that $\better(\bfx,\theta) \subseteq \setKper_{\langle \bar{\theta},k \rangle}$, from (\ref{eq_defbettermap}) it is sufficient to verify the following for all $j\in \mathcal{N}$ and $\bfy\in \better_j(\bfx,\theta)$:
\begin{subequations}
	\label{eq:cond_y}
	\begin{equation}
	\label{eq:max_y}
	\bfy_i \in [0,\mu_i]^{T|\setP|} \qquad \forall i\in \mathcal{N}.
	\end{equation}
	\begin{equation}
	\label{eq:totsum_y}
	\sum_{\tau\in{\mathcal{T}}} \sum_{p\in \setP_i} \bfy_{i\{\tau\} p}  = D_i \qquad \forall i\in \mathcal{N}.
	\end{equation}
\end{subequations}
Both conditions are implied by (\ref{eq:cond_x}) as a result of expressions (\ref{eq_deltamax}) and (\ref{eq:mapping_eq1}), respectively.

\textit{Assumption \ref{assum_rhmap}-\eqref{assum_rhmap_c}}:
From the nonnegativity of the last two terms in (\ref{eq:lambda}), it is straightforward to verify the following conditions on $\Delta(\bfx,s_i)$ for any $i \in \mathcal{N}$ and $s_i\in \setS_i(\bfx,\bar{\theta})$:
\begin{subequations}
\label{eq:delta_cond}
\begin{equation}
\label{eq:delta_cond1}
\Delta(\bfx,s_i) = 0 \qquad \text{if } \lambda(\bfx,s_i)\leq0
\end{equation}
\begin{equation}
\label{eq:delta_cond2}
\Delta(\bfx,s_i) > 0 \qquad \text{if } \lambda(\bfx,s_i) > 0
\end{equation}
\end{subequations}
Moreover, given expression (\ref{eq:mapping_eq1}), it holds:
\begin{subequations}
\label{eq:M_cond}
\begin{equation}
\label{eq:M_cond1}
\better_i(\bfx,\theta)=\{x  \}  \qquad \text{if } \Delta(\bfx,s_i) = 0
\end{equation}
\begin{equation}
\label{eq:M_cond2}
\better_i(\bfx,\theta) \cap \{ x \} = \varnothing \qquad \text{if } \Delta(\bfx,s_i) > 0
\end{equation}
\end{subequations}
From the above results, since the mapping $\better$ is given by the composition (\ref{eq_defbettermap}), it is sufficient to verify the following to conclude the proof:
\begin{subequations}
\label{eq:Vineq}
\begin{equation}
\label{eq:Vineq1}
V \left ( F(\bfx,s_i) \right) - V(\bfx) = 0 \qquad \text{if } \Delta(\bfx,s_i)=0
\end{equation}
\begin{equation}
\label{eq:Vineq2}
V \left ( F(\bfx,s_i) \right) - V(\bfx) < 0 \qquad \text{if } \Delta(\bfx,s_i)>0
\end{equation}
\end{subequations}
Since (\ref{eq:Vineq1}) straightly follows from (\ref{eq:M_cond1}), we now demonstrate condition (\ref{eq:Vineq2}) by evaluating its left-hand side when $\Delta(\bfx,s_i)>0$. This can be written as:
\begin{multline}
\label{eq:Vest}
V \left ( F(\bfx,s_i) \right) - V(\bfx) \overset{(\ref{eq:Vest}a)}{=} \sum_{l\in \pa} \Gamma \left( \Lambda_l (\bfx_{\{ \ta \}}) + \Delta(\bfx,s_i)  \right)  -  \Gamma \left( \Lambda_l (\bfx_{\{ \ta \}})  \right)\\
+  \sum_{l\in \pda} \Gamma \left( \Lambda_l (\bfx_{\{ \tda \}}) - \Delta(\bfx,s_i)  \right)  -  \Gamma \left( \Lambda_l (\bfx_{\{ \tda \}})  \right) \\
\overset{(\ref{eq:Vest}b)}{=} \sum_{l\in \pa} \int_0^{\Delta(\bfx,s_i)} \Pi \left( \Lambda_l (\bfx_{\{ \ta \}}) + \sigma  \right) \, d\sigma -  \sum_{l\in \pda} \int_0^{\Delta(\bfx,s_i)} \Pi \left( \Lambda_l (\bfx_{\{ \tda \}}) - \sigma  \right) \, d\sigma  \\
\overset{(\ref{eq:Vest}c)}{\leq} \sum_{l\in \pa} \int_0^{\Delta(\bfx,s_i)} \Pi \left (  \Lambda_l (\bfx_{\{ \ta \}}) \right )  + \kappa_{\Pi} \sigma   \, d\sigma - \sum_{l\in \pda} \int_0^{\Delta(\bfx,s_i)} \Pi \left( \Lambda_l (\bfx_{\{ \tda \}})  \right) - \kappa_{\Pi} \sigma \, d\sigma \\
=\sum_{l\in \pa} \left ( \Pi \left( \Lambda_l (\bfx_{\{ \ta \}})  \right) + \frac{\Delta(\bfx,s_i)}{2} \kappa_{\Pi}\right ) - \sum_{l\in \pda} \left( \Pi \left( \Lambda_l (\bfx_{\{ \tda \}})  \right) - \frac{\Delta(\bfx,s_i)}{2} \kappa_{\Pi}  \right)\\
= \left [ \sum_{l\in \pa}\Pi \left( \Lambda_l (\bfx_{\{ \ta \}})  \right) - \sum_{l\in \pda}\Pi \left( \Lambda_l (\bfx_{\{ \tda \}})  \right) \right ] + \frac{\Delta(\bfx,s_i)}{2} \left ( |\pa| + |\pda| \right)
\end{multline}
where (\ref{eq:Vest}a) and (\ref{eq:Vest}b) derive from expression (\ref{eq:mapping_eq1}) and (\ref{eq:Gamma}), respectively, whereas (\ref{eq:Vest}c) follows from Assumption \ref{assum:Lipschitz}. To check (\ref{eq:Vineq2}) and conclude the proof, it is sufficient to verify that the last term in (\ref{eq:Vest}) is always negative from (\ref{eq:pi_p}) and (\ref{eq_deltamax}) as we are assuming $\Delta(\bfx,s_i)>0$.

\textit{Assumption \ref{assum_rhmap}-\eqref{assum_rhmap_d}}: From (\ref{eq:M_cond}), only the first condition in (\ref{eq:rh_map_d}) needs to be verified.
This is done by contradiction: if the considered $\bfx_i$ is not a minimizer of $J_i$, player
$i$ can swap part of its transmission from some path $\pda$ at time $\tda$ to some other path $\pa$ at time $\ta$. Equivalently, there must exist $s^*_i=(\ta,\tda,\pa,\pda)\in \setA(\tcapi_i) \times \setP_i  \times \setP_i$ such that:
\begin{equation}\label{eq:ioptcond}
\left\{ \begin{array}{l}
	\pi_{\pda} ( \bfx_{\{\tda\}}) - \pi_{\pda} ( \bfx_{\{\tda\}})  > 0,\\
	\mu_i - \bfx_{i\{\ta\}\pa}> 0,\\
	\bfx_{i\{\tda\}\pda} > 0.
\end{array}
\right.
\end{equation}
Since $\lambda(\bfx,s_i^*)$ in (\ref{eq:lambda}) is positive, from (\ref{eq_Sn}), we have that $\lambda(\bfx,s_i)\geq \lambda(\bfx,s_i^*) >0 $  for any $s_i \in \setS_i(\bfx,\tcapi_i)$.
The result is then verified from (\ref{eq:delta_cond2}) and (\ref{eq:Vineq2}).

\bibliographystyle{apalike}
\bibliography{biblioDR}

\end{document}